\documentclass[useAMS,usenatbib]{mn2e}
\usepackage{eso-pic,graphicx}
\usepackage{float}
\usepackage[caption = false]{subfig}

\usepackage{color,multirow}

\usepackage[backref,breaklinks,colorlinks,citecolor=blue]{hyperref}
\usepackage[all]{hypcap}
\usepackage{amsmath}
\newcommand{\angstrom}{\textup{\AA}}

\title[Extinction towards Westerlund~1] {Extinction law in the range 0.4 - 4.8 $\mu$m and the 8620\,\AA\,DIB towards the stellar cluster Westerlund~1 \thanks{Based on observations obtained at the  Cerro Tololo Inter-American Observatory (CTIO),  Observat\'orio do Pico dos Dias (OPD/LNA/MCT),  Southern Astrophysical Research (SOAR), European Southern Observatory (ESO) and  Gemini South Observatory (GS).}}
\author[Damineli et al.]
  {A. Damineli$^{1}$\thanks{E-mail: augusto.damineli@gmail.com},
  L.~A.~Almeida$^{1}$, 
  R.~D.~Blum$^{2}$,
  D.~S.~C.~Damineli$^{3,4}$,
  F.~Navarete$^{1}$, \\
  \newauthor
  M.~S.~Rubinho$^{1}$,
  M.~Teodoro$^{5,6}$
   \\
  $^{1}$~Instituto de Astronomia, Geof\'isica e Ci\^encias Atmosf\'ericas da USP\\
   Rua do Mat\~ao 1226, Cidade Universit\'aria S\~ao Paulo-SP, 05508-090, Brasil \\
  $^{2}$NOAO, 950 N Cherry Ave., Tucson, AZ 85719 USA \\
  $^{3}$Cell Biology and Molecular Genetics Department, University of Maryland, College Park, Maryland 20742-5815, USA \\
  $^{4}${~PhD Program in Computational Biology, Instituto Gulbenkian de Ci\^encia, 2780-156 Oeiras, Portugal}\\
  $^{5}${~Astrophysics Science Division, Code 667, NASA Goddard Space Flight Center, Greenbelt, MD 20771, USA} \\
  $^{6}${~Universities Space Research Association, 7178 Columbia Gateway Dr., Columbia, MD 20146, USA}\\
  }
\date{}

\def\LaTeX{L\kern-.36em\raise.3ex\hbox{a}\kern-.15em
    T\kern-.1667em\lower.7ex\hbox{E}\kern-.125emX}

\begin{document}

\label{firstpage}

\maketitle

\begin{abstract}
The young stellar cluster Westerlund~1 (Wd~1: $l$\,=\,339.6$^\circ$, b\,=\,$-$0.4$^\circ$) is one of the most massive in the local Universe, but accurate parameters are pending on better determination of its extinction and distance. Based on our photometry and data collected from other sources, we have derived a reddening law for the cluster line--of--sight representative of the Galactic Plane (-5$^\circ<$\,b\,$<$+5$^\circ$) in the window 0.4-4.8\,$\mu$m: The power law exponent $\alpha$\,=\,2.13$\pm$0.08 is much steeper than those published a decade ago (1.6--1.8) and our index $R_V$\,=\,2.50\,$\pm$\,0.04 also differs from them, but in very good agreement with recent works based on deep surveys in the inner Galaxy. As a consequence, the total extinction $A_{Ks}$\,=\,0.74$ \pm $0.08\ ($A_V$\,=\,11.40$ \pm$ 2.40) is substantially smaller than previous results(0.91--1.13), part of which ($A_{Ks}$\,=\,0.63 or $A_V$\,=\,9.66) is from the ISM.
 The extinction in front of the cluster spans a range of $\Delta A_V\sim$8.7 with a gradient increasing from SW to NE across the cluster face, following the same general trend of warm dust distribution. The map of the $J-Ks$ colour index  also shows a trend of reddening in this direction. We measured the equivalent width of the diffuse interstellar band at 8620~\AA\ (the ``GAIA DIB'') for Wd~1 cluster members and derived the relation $A_{Ks}$\,=\,0.612\,$EW$ $-$ 0.191\,$EW^2$. This extends the \citet{Munari+2008} relation, valid for $E_{B-V}$\,$<$\,1, to the non--linear regime ($A_V$\,$>$\,4).

\end{abstract}

\begin{keywords}
  Galaxy: open clusters and associations: individual: Westerlund 1 - ISM: general - ISM: lines and bands - ISM: dust, extinction: 
\end{keywords}

\section{Introduction}\label{section1}
The majority of stars are born in large clusters, and so, these sites are key to understanding the stellar contents of galaxies. Until recently the Milky Way was believed to be devoid of large young clusters. The realisation that our Galaxy harbours many young clusters with masses (M\,$>$\,10$^3$\,M$_{\odot}$) like Westerlund~1 (Wd~1), Arches, Quintuplet, RSG~1, RSG~3, Stephenson~2, Mercer~81, NGC3603, h+$\chi$ Persei, Trumpler~14, Cygnus~OB2, [DBS2003] and VdBH~222 \citep[see summary by][]{Negueruela2014}, reveals a different scenario than previously thought. 
Although Milky Way offers the opportunity to resolve the cluster members, unlike other galaxies beyound the Local Group, accurate determination of the fundamental properties of these recently discovered clusters: distance, mass, age, initial mass function (IMF), and binary fraction is still lacking in many cases.  The two fundamental parameters upon which all others depend are the interstellar extinction and the distance. In the present work we use accurate techniques to derive the interstellar extinction towards Wd~1 and its surroundings, what was not accurately done in previous works.

The extinction is related to the observed magnitudes by the fundamental relation (distance modulus):
\begin{equation}
m_{\lambda} = M_{\lambda} + 5\log_{10} \left(\frac{d}{10}\right) + A_{\lambda}.
\label{eq1}
\end{equation}

A set of different filters can be combined to define colour excess indices: $\displaystyle E_{\lambda1}-E_{\lambda2}=(m_{\lambda1}-m_{\lambda2})_{\rm{obs}}-(m_{\lambda1}-m_{\lambda2})_0$, where the zero index indicates the intrinsic colour of the star. Some authors -- like the classical work of \citet{Indebetouw+2005} -- attempted to derive $\displaystyle A_\lambda$ directly from the above relations, using a minimization procedure to a large number of observations from 2MASS survey \citep{Stru06}. Actually the number of variables is greater than the degrees of freedom of the system of equations we have to compute.  All possible colour excess relations are linear combinations of the same parameters. In the specific case of the NIR, after dividing by the K$_s$--band extinction, there will be an infinite number of pairs $\displaystyle A_J/A_{Ks}$ and  $\displaystyle A_H/A_{Ks}$ satisfaying the relations. Many minimization programs (like the downhill technique used in the {\it amoeba} program) just pick up a local minimum close to the first guess as a solution, hiding the existence of other possibilities. Derivation of the extinction can only be accomplished on the basis of a specific extinction law as a function of wavelength, which ultimately reflects the expected properties of dust grains.

Interstellar reddening is caused by scattering and refraction of light by dust grains, and the amount of light subtracted from the incoming beam (extinction) is governed by ratio between its wavelength ($\lambda$) and the size of the grains (d). For $\lambda<<d$ all photons are scattered/absorbed (gray extinction).For $\lambda>d$ the fraction of photons that escape being scattered/absorbed increases. The interstelar dust is a mixture of grains of different sizes and refactory indices, leading to a picture a little more complicated than described above. This was first  modelled by \citet{Hulst1946} for different dust grain mixtures. All subsequent observational works resulted in optical and NIR extinction laws similar to that of van de Hulst model (in particular his models \#15 and \#16). A remarkable feature is that they are well represented by a power low ($A_\lambda$\,$\propto$\,$\lambda^{-\alpha}$) in the range 0.8\,$<$\,$\lambda$\,$<$\,2.4\,$\mu$m \citep[see e.g.][]{F99}. 

The $\alpha$ exponent of the extinction power law: $\displaystyle A_{\lambda}/A_{Ks} = (\lambda_{Ks}/\lambda)^\alpha$ is related to the observed colour excess through:
	 
\begin{equation}
\frac{A_{\lambda_1}-A_{\lambda_2}}{A_{\lambda_2}-A_{\lambda_{Ks}}} = \frac{\left(\frac{\lambda_2}{\lambda_1}\right)^{\alpha}-1}{1-{\left(\frac{\lambda_{Ks}}{\lambda2}\right)^{\alpha}}}. 
\label{eq2}
\end{equation}

The value of the $\alpha$ exponent is driven by: a) the specific wavelength range covered by the data; b) the effective wavelengths of the filter set, which may differ from one to another photometric system, especially the $R$ and $I$ bands; and c) the fact that the effective wavelength depends on the spectral energy distribution (SED) of the star, the transmission curve of the filter and the amount of reddening of the interstellar medium (ISM). The power law exponent {\large $\alpha$} in the range 1.65\,$<$\,$\alpha$\,$<$\,2.52 has been reported \citep[see e.g.,][]{CCM89, Berdnikov+1996, Indebetouw+2005, Stead+2009, RL85, FM09, Nishiyama+2006, GF14}, but it is not clear how much spread in the value of the exponent is due to real physical differences in the dust along different lines-of-sight and how much comes from the method used on the determination of the exponent. As shown by \citet[their Fig. 5]{GF14} using the 2MASS survey, the ratio of colour excess $E_(H-K)/E_(J-K)$ grows continuously from 0.615 to 0.66 as the distance modulus grows from 6 to 12 towards the inner Galactic Plane (GP, their Fig. 5). This corresponds to a change in $\alpha$ from 1.6 to 2.2 which translates into $A_J/A_{Ks}$ from 2.4 to 3.4. \citet{Zas09} also used 2MASS data to show that colour excess ratios varies as a function of Galactic longitude, indicating increasing proportions of smaller dust grains towards the inner GP. Reddening laws steeper than the''canonical” ones have been suggested for a long time, but their reality is now clearly evident from deep imaging surveys. 

The large progress reached in recent years revealed that there is no ``universal reddening law'', as believed in the past. Moreover, the extinction coefficients are quite variable from one line--of--sight to another, even on scales as small as  arcminutes. At optical bands ($UBV$) it was already well established long ago that the extinction law for particular lines-of-sight had large discrepancies from the average \citep{FM09,He+1995,Popowski2000,Sumi2004,Racca+2002}. In recent years this has been proved to occur also for the NIR wavelength range \citep{Larson+2005, Nishiyama+2006, Froebrich+2007, Gosling+2009}. The patchy tapestry of extinction indices is particularly impressive in the large area work in the NIR$/$optical done by \citet{Nataf16,Nataf13} for the Galactic Bulge and by \citet{Scha16} for the GP. Although we cannot use directly the extinction coefficients from these works for our particular target (line--of--sight), their derived reddening relations help for checking the consistency of our results. Targets to derive the extinction must be selected between stars with well known intrinsic colour indices, in order to measure accurate colour excesses. Wd~1 cluster members with known spectral types are ideal for this, especially because the majority of them are hot stars, for which the intrinsic colours are close to zero. There are $\approx$ 92 stars in this group; the statistics can be improved by using stars in the field around the cluster.

\citet{Wozniak+1996} proposed using Red Clump (RC) stars, to derive the ratio between the total to selective extinction. These Horizontal Branch stars are the most abundant type of luminous stars in the Galaxy and have a relatively narrow range in absolute colours and magnitudes. RCs form a compact group in the colour-magnitude diagram (CMD) as shown by \citet[and references therein]{Stanek+2000}. This is due to the low dispersion -- a few tenths of magnitude -- in intrinsic colours and luminosities of RC stars  \citep{Stanek+1997,Paczynski+1998}. This technique, initially designed for using filters $V$ and $I$ in OGLE survey for microlensing events, and filters $V$ and $R$ in MACHO, was adapted for the $JHKs$ bands \citep{Flaherty+2007,Indebetouw+2005,Nishiyama+2006,Nishiyama+2009}. 

As shown by \citet{Indebetouw+2005} and by \citet{Nishiyama+2006,Nishiyama+2009}, RC stars in the CMD (e.g. $J-Ks$ {\it versus} $Ks$) may appear as an over--density ``strip''. That strip in the CMD contains interlopers which mimic RC star colours but have different luminosities (like nearby red dwarfs and distant supergiants). This does not allow the application of the relation~\eqref{eq1} to derive the absolute extinction from each particular star in the strip, but still works for the colour excess ratios in the relation~\eqref{eq2}. From the measured colour excess ratio (e.g. $E_{J-H}/E_{J-Ks}$) the value of the exponent $\alpha$ can be calculated and, therefore, the ratios $A_J/A_{Ks}$ and $A_H/A_{Ks}$.

\citet{Nishiyama+2006} reported $\alpha$\,=\,1.99 and $A_J$\,$\approx$\,3.02 and $A_H$\,$\approx$\,1.73 in a study of the Galactic Bulge, which is much higher then all previous results. \citet{Fritz+2011} also derived a large value $\alpha$\,=\,2.11 for the Galactic Centre (GC), using a completely different technique. \citet{Stead+2009} reported $\alpha$\,=\,2.14\,$\pm$\,0.05 from UKIDSS data and similar high exponents from 2MASS data. They did not derive $A_\lambda/A_{Ks}$, since in their approach those quantities vary because of shifts in the effective wavelengths as the extinction increases (see Section 2.2). However, at a first approximation, using the isophotal wavelengths, we can calculate from their extinction law: $A_J/A_{Ks}$\,$\approx$\,3.25 and $A_H/A_{Ks}$\,$\approx$\,1.78. 

The dream of using interstellar DIBs to evaluate the exinction has been hampered by saturation effects in the strength of the features and on the behaviour of the carriers which differ from the hot/diffuse ISM as compared to cold/dense clouds -- however, see \citet{Maiz+2015}. The 8620\,\AA\,DIB correlates linearly with the dust extinction \citep{Munari+2008}, at least for low reddening, and is relatively insensitive to the ISM characteristics. Since this spectral region will be observed by GAIA for a large number of stars up to relatively large extinction, we used our data to extend \citet{Munari+2008} relation, which was derived for low reddening.

This work is organised as follows. In Section~\ref{section2} we describe the photometric and spectroscopic observations and data reduction. In Section~\ref{section3} we describe the colour excess ratios relations, the ratio between the absolute magnitudes and a suggested extinction law for the inner GP.  In Section~\ref{section4} we compare our results with others reported in the literature. In Section~\ref{section5} we perform $J-{Ks}$ extinction maps around Wd~1 field for a series of colour slices and evaluate the 3D position of obscuring clouds. In Section~\ref{section6} we analyse the relation between the interstellar extinction and the Equivalent width (EW) of the 8620~\AA\ DIB. In  Section~\ref{section7} we present our conclusions.

\section{Observations and data reduction}\label{section2}

\begin{figure}
 \centering
 \resizebox{\hsize}{!}{\includegraphics[width=20.0cm,angle=0, trim= 0cm 0.7cm 0cm 2cm, clip]{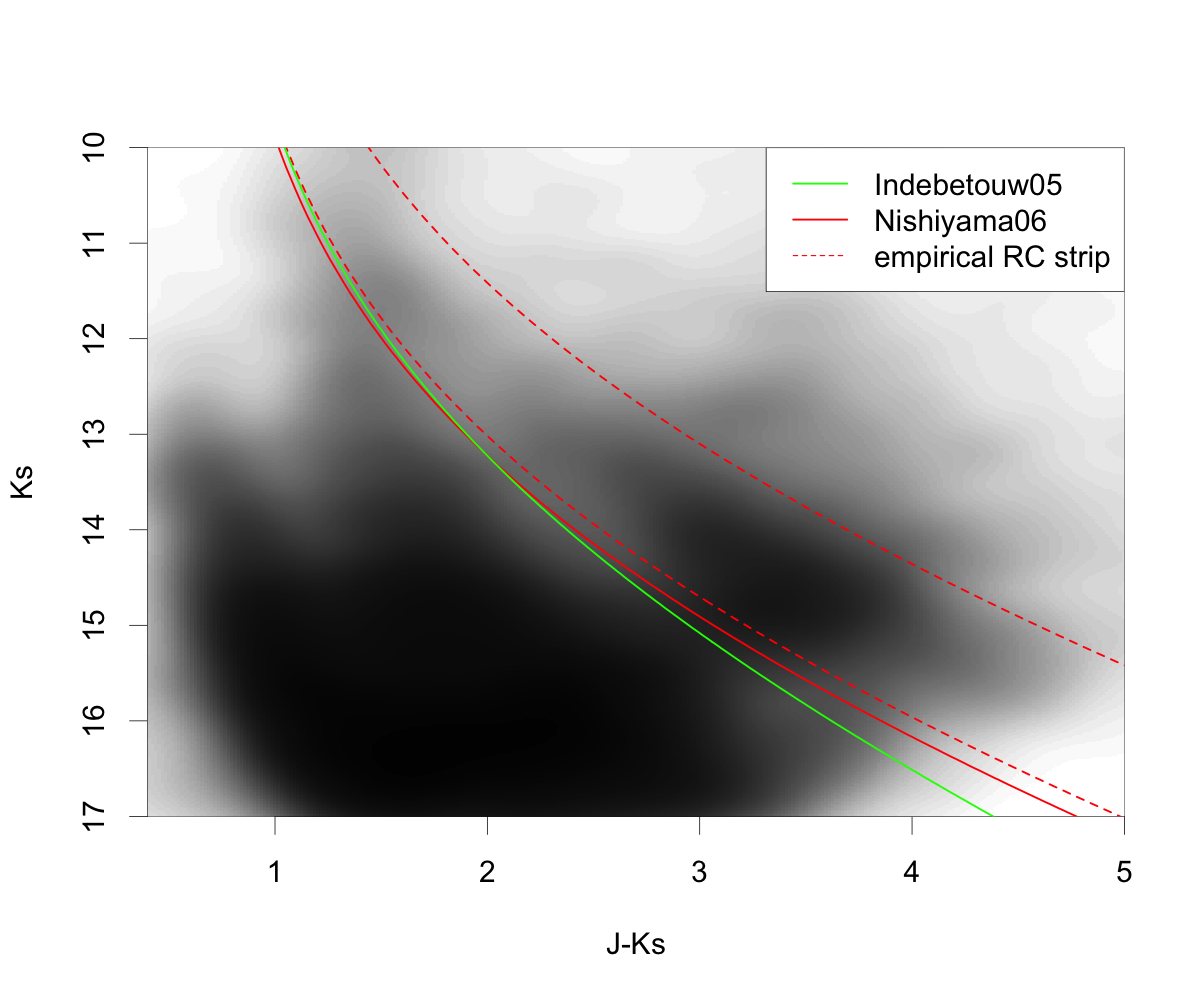}}
 \caption{The $J-Ks$ vs. $Ks$ Hess diagram for Wd~1 field showing the RC strip. The two lower continuous lines correspond to extinction laws by \citet{Indebetouw+2005} (blue) - with extinction modules $\mu_{Ks}$\,=\,0.15\,mag\,kpc$^{-1}$ - and by \citet{Nishiyama+2006} (red) - with $\mu_{Ks}$\,=\,0.1\,mag\,kpc$^{-1}$. The two dashed red linear are empirical limits containing the RC strip, using Nishiyama's law with lower limit $\mu_{Ks}$\,=\,0.11 and upper limit 0.23 mag\,kpc$^{-1}$.}
 \label{fig:RCstrip}
\end{figure}

\subsection{Photometry}\label{section2.1}
The main set of images in the $JHKs$ filters was taken on 2006 June 06 with the ISPI camera at the 4m Blanco Telescope (CTIO), with 10\arcmin\,$\times$\,10\arcmin\ FOV \citep{Bliek+2004}. The image quality, after combination of the dithered sub-images was 1.0\arcsec in the $Ks$ filter and 1.8\arcsec in $J$ and $H$.  The number of stars with measurements in all the three filters was 23834 with errors less than 0.5\,mag. The limiting magnitudes were: $J$\,$<$\,21.4, $H$\,$<$\,19.2 and $Ks$\,$<$\,18.2. The basic calibration (flatfield, linearity correction) was done with {\tt IRAF}\footnote{http://www.iraf.noao.edu} and the photometry extraction was performed with \texttt{STARFINDER}\footnote{http://www.bo.astro.it/StarFinder/paper6.htm}.

In order to cover the brighter stars, we took $JHKs$ imaging at the 1.6-m and 0.6-m OPD/LNA Brazilian telescopes with the CAMIV camera. Many images were taken in several subsequent years, covering a FOV larger than that of ISPI. For each star, we always use the three $JHKs$  magnitudes taken in a single night, to preserve the stellar colours against variability. In addition to traditional imaging, we used different strategies to record the very bright stars, for example, deploying a 5\,mag neutral density filter or masks with four 5\,cm diameter holes in front of the telescope. We compared the results obtained with the different setups and found good agreement. Instead of the $Ks$ filter, we used a narrow filter with central wavelength 2.14\,$\mu$m (FWHM\,=\,0.02\,$\mu$m) and cross calibrated the photometry with $Ks$ magnitudes from the ISPI camera. The uncertainty on our $JHKs$  magnitudes reached at most 0.1\,mag for a few bright stars like W26, when comparing magnitudes at different epochs, which could be due to intrinsic variability. 

Data processing was made in the same way as with ISPI images. Our catalogue in $JHKs$ from these two cameras is $\approx$\,100\% complete for the blue and red giants and blue supergiants/hypergiants down to the bottom of the Main Sequence (MS, $Ks$\,$\sim$\,15). The peak of the luminosity function in the $Ks$ filter (completeness $>$\,85\%) is at $Ks$\,=\,16.0). We call this set of data the ISPI+CAMIV photometric sample.

The calibration of ISPI photometry was performed against {\tt 2MASS}\footnote{http://irsa.ipac.caltech.edu/cgi-bin/Gator/}. Then, we used the CTIO/ISPI catalogue and stars in common with OPD/CAMIV photometry performed at smaller telescopes to extend 2MASS calibrations up to $Ks$\,$\approx$\,0. The central 2\arcmin\,$\times$\,2\arcmin field of Wd~1 is so crowded that some stars were resolved only in nights with excellent seeing. The Spartan camera at the SOAR telescope was used for this purpose on 2013 September 03 when the seeing was approximately 0.5\arcsec.

We used the VVV survey for $ZY$ magnitudes of the RC candidates we detected with ISPI and CAMIV cameras. The magnitudes were extracted using aperture photometry from the database of the survey\footnote{http://horus.roe.ac.uk/vsa/}. The date of those images was 2010 July 19th. For those specific stars, we used our ISPI JHKs magnitudes (calibrated in 2MASS system) to transform from VISTA magnitdes to 2MASS. However, even if the $JHKs$ magnitudes obtained with VVV survey were in excellent agreement with our ISPI+CAMIV+Spartan catalogue, we used them for two specific aims: a) to perform the colour index density maps and b) for checking the correct identification of the ISPI/CAMIV sources. We used magnitudes for the W1 and W2 MIR bands of the WISE survey \citep{wright10}. We used also images from Herschel Hi--GAL survey to look for cold and warm dust \citep{molinari10}.

For the Wd~1 cluster members, we used $BVI$ photometry from \citet{Lim+2013}. A few stars without $B$ or $V$ magnitudes from those authors were taken from \citet[]{Clark+2005}, who kindly made a machine readable copy of the photometry available to us. We recalibrated those magnitudes to \citet{Lim+2013} system. Additional magnitudes were taken from \citet{Bonanos2007} and also transformed to the \citet{Lim+2013} system. For the $R$ filter, we used the data from \citet{Clark+2005} and \citet{Bonanos2007} without any transformation. We used 104 stars with spectral classification done by \citet{Clark+2005}, \citet{Negueruela+2010} and \citet{Ritchie+2009}, which have accurate intrinsic colour indices. For reddening studies, we included also the Main Sequence OB eclipsing binary Wddeb \citep{Bonanos2007}. The majority of the bright cluster members are OB Supergiants, and we included also cooler giants and hypergiants. We excluded  W8a, W9, W12a,  W20, W26, W27, W42a, W75, W265 and the WC9 WRds which have circumstellar  hot dust emission (in the $Ks$ filter) or absorption by dust (in $B$ filter). 

\begin{figure}
\centering
\resizebox{\hsize}{!}{\includegraphics[width=20.0cm,angle=0, trim= 0cm 0.65cm 0cm 2cm, clip]{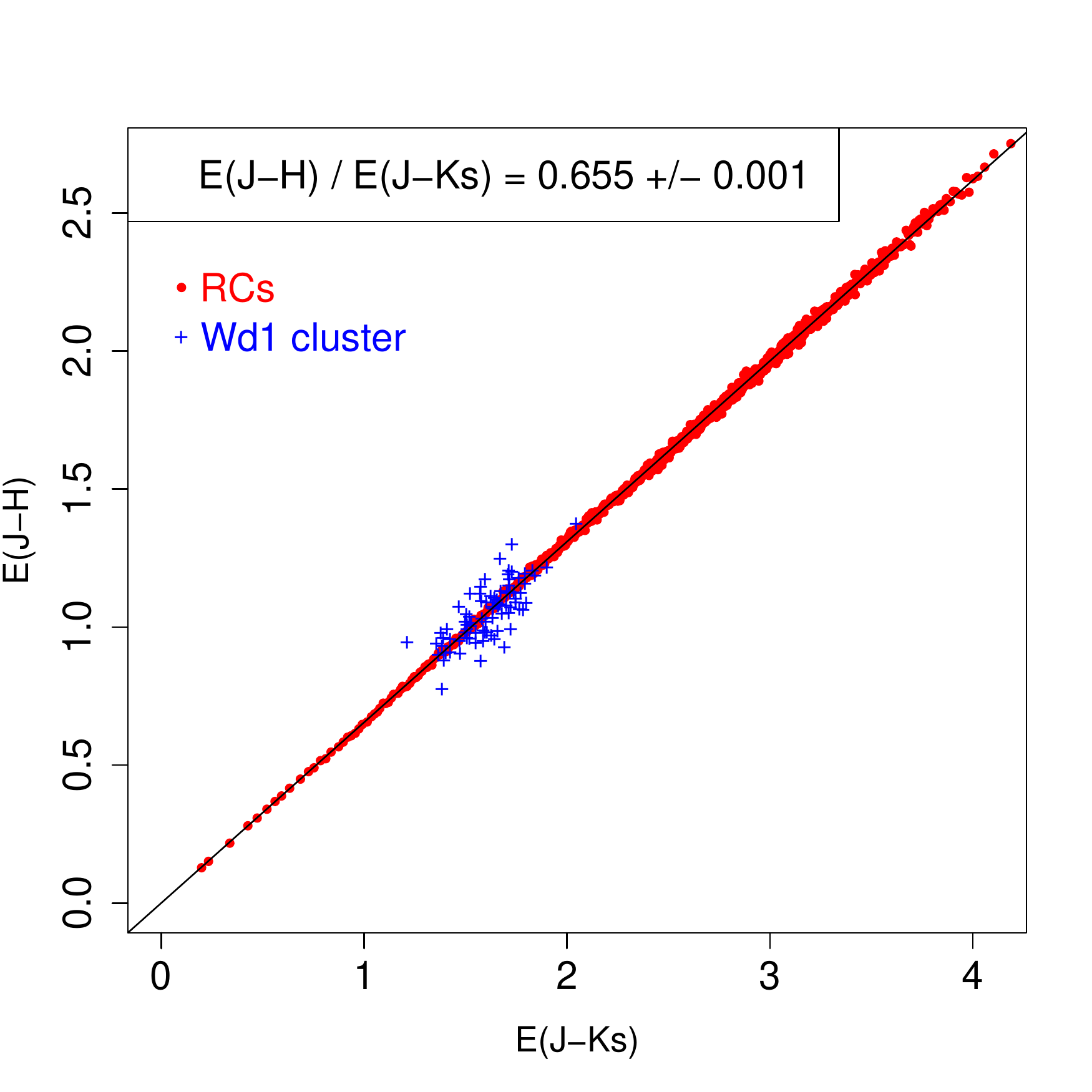}}
\caption{The excess colour ratio is $E_{J-H}/E_{J-Ks} = 0.655 \pm 0.001$. RCs are in red points (averaged in groups of 10) and Wd~1 cluster members are blue crosses and were used together RCs in the linear fit. It is clear that there is a differential extinction among the Wd~1 cluster members and their average value is intermediate as compared to the RC sample. See other plots in Appendix B.}
\label{fig:EJHXEJK}
\end{figure}

There have been discussions about the accuracy of the optical data by the cited authors, since the colour indices of the program stars were larger than that of photometric standards. Errors could amount up to a few tenths of magnitudes in the B filter and a little less at longer wavelengths. This has an impact when comparing magnitudes from different authors, but has a negligeable effect for other purposes, like study of variability in time series taken with the same instrument. 
The impact is minor regarding the study of reddening due to the fact that the extinction is orders of magnitudes larger than photometric errors. Similar systematic errors due to the lack of very red photometric standards is a common place in large deep surveys, and, although they are present at all wavelength windows, they have not been discussed in connection to deep surveys.  

We used extensively the R\footnote{https://www.R-project.org/} programming packages to perform statistics, fitting and plots.

\subsection{Effective filter wavelengths}\label{section2.2}

Since JHKs filters are close in wavelengths, special care must be taken, using accurate effective wavelengths for this particular range. As a matter of fact every single star is measured at different wavelength, defined by the intrinsic SED of the star, modified by its particular amount of reddening and convolved with the filter passband. This explains why the colour--colour diagram (CCD) produces curved bands. The effective wavelength of the filters shifts continuously to longer wavelengths as the reddening increases, such that the ratios vary continuously. The shifts in the effective wavelengths are approximately straight lines (see Fig.~2a in \citet{Stead+2009}), in order that ratios between effective wavelengths are reasonably constant for large extinctions. Linear approximations for the effective wavelengths of 2MASS filters taken from that plot are presented in Apendix\,\ref{appendixA}. 

Although a colour plot $J-$H\,{\it vs.}\,$H-Ks$ does not produce a straight reddening line, the curvature of the line is small for large reddening and the ratio converges to a constant value. In this way, we could use a template star SED for evaluating the effective filter wavelengths.  An approach simpler to the one we discussed in the last paragraph was also used by \citet{Stead+2009}, who convolved the 2MASS filter passbands with a \citet{Castelli+2004} K2III giant star spectrum, obtaining  $\lambda_J$\,=\,1.244, $\lambda_H$\,=\,1.651 and $\lambda_{Ks}$\,=\,2.159. Since our data are from deep imaging towards the inner Galaxy and are calibrated in the 2MASS system, we use the above effective wavelengths for $JHKs$ filters (Table\,\ref{table1}, column~2). 

Other filter wavelengths were taken as they are repported in the parent papers and not shifted, as we do not use them for critical calculations. Care must be taken because the same filter names have different wavelengths in different papers, especially $R$, $I$, $Z$ filters. For example, the $R$ filter can have $\lambda_{\rm eff} = 0.664$ or  0.70 $\micron$. For the $I$ filter, the most used in recent works is $\lambda_{\rm eff}$\,=\,0.805\,$\mu$m, but  \citet{CCM89} uses 0.90\,$\mu$m, which in reality matches better the $Z$ band. To avoid confusion, we state in Table\,\ref{table1} the effective wavelength of each filter we are using. 

\subsection{Spectroscopy}\label{section2.3}

For the 8620 \AA DIB, we observed 11 bright members from the Wd~1 cluster and 12 along other Galactic directions at the Coud\'e focus of the 1.6-m telescope. The OPD/LNA spectra have $R$\,=\,15000. In addition, we downloaded 31 spectra from the ESO database (see Table\,\ref{table2}) also with $R$\,$\sim$\,15000. OPD data were reduced with IRAF; the ESO prog. ID 073D-0327 data with ESO Reflex and ESO 081D-0324 with Gasgano. The spectral resolution was measured on telluric absorption lines. When there were stars in both samples, we used the OPD/LNA because its Deep Depletion CCD presented no fringes, which are prominent in some ESO spectra, especially to the red of the 8620\,DIB. EWs were measured by Gaussian fitting after the spectra were normalised to the stellar continuum. Typical uncertainty is $\sim$\,10\% for the bright and blue stars, but errors are difficult to asses for fainter stars. EWs were also difficult to measure in spectral types later than F5, because of blends with stellar lines. Our results are presented in Table\,\ref{table2} column~2. One specific star, the eclipsing binary Wddeb was measured with Gemini South/GMOS with spectral resolution $R$\,$\sim$8000.

\section{The reddening law using colour excesses }\label{section3}
When the intrinsic colours of a star are known, its colour excesses can be derived from the observed colours, and the reddening law is calculated from the ratios between colour excesses. This is the case for the Wd~1 cluster members with spectral classification reported in \citet{Clark+2005}, \citet{Negueruela+2010} and \citet{Ritchie+2009}. The intrinsic colours were derived using calibrations by  \citet{Wegner1994}. For the Wolf-Rayet stars, we used \citet{Crowther+2006}. Since the number of Wd~1 cluster members are not large, additional stars were taken from the surrounding field. They do not have spectral classification, but for RC stars there are reliable intrinsic colour indices (see below).

\subsection{Selection of Red Clump stars}\label{section3.1}
A similar procedure to that of \citet{Wozniak+1996} was adopted for selecting RC candidates as they can easily be identified as an overdensity in the CMD (see Fig.\,\ref{fig:RCstrip}). We represented the photometry by a Hess diagram in order to see more clearly the overdensity structures. The lower left overdensity corresponds to foreground (low luminosity) red stars. The one that goes up vertically, around $J-Ks \approx 1.5$ is the Main Sequence (MS) of the Wd~1 cluster and the large region which curves to the red below $Ks\approx 14$ is the cluster Pre-Main Sequence (PMS). The overdensity region running brighter/bluer to fainter/redder in the middle of the CMD is the expected locus containing RC stars observed at different distances and amounts of extinction. For a preliminary study, we plotted in our observed $Ks$\,$\times$\,$J-Ks$ CMD the position of a typical RC star by varying its distance and using different extinction laws like those of \citet{Indebetouw+2005}, who reported  a ``extinction modulus'' $\mu_{Ks}$\,=\,$A_{Ks}/d$\,=\,0.15\,mag\,kpc$^{-1}$ - blue line. 
We followed the same procedure using the reddening law of \citet{Nishiyama+2006}, who do not report the ``extinction modulus'', but which can be derived from the RC peak in their Fig.\,2, corresponding to $\mu_{Ks}$\,=\,0.1\,mag\,kpc$^{-1}$.  Although we do not use the extinction modulus for any particular measurement (since it varies from star to star) Fig.\,\ref{fig:RCstrip} indicates that typical extinction in the direction of Wd~1 cluster is much larger than for directions explored by \citet{Indebetouw+2005} - large Galactic longitudes, or by \citet{Nishiyama+2006} - the Bulge.

\begin{figure}
 \centering
 \resizebox{\hsize}{!}{\includegraphics[width=20.0cm,angle=0, trim= 0cm 0.5cm 0.5cm 1.5cm, clip]{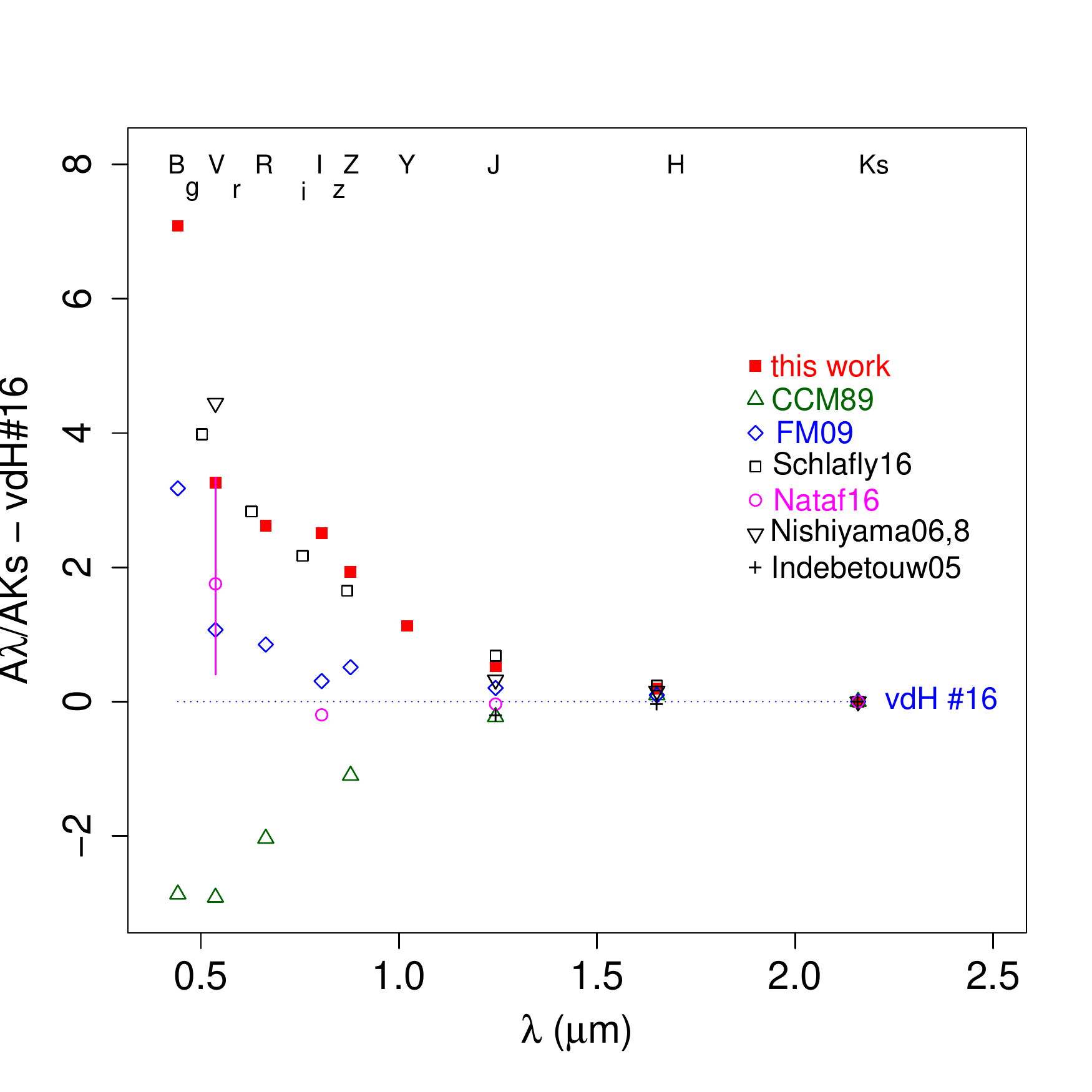}}
 \caption{Differences between $A_\lambda/A_{Ks}$ for a set of reddening laws and that of \citet{Hulst1946}, taken as a zero point.The vertical magenta line is the range reported by \citet{Nataf16}. Data assigned as Schlafly16 \citep{Scha16} in reality were derived by us, using their colour excess ratios. The letters at the top of the plot indicate the approximate wavelengths of the filters - see Table\,\ref{table1}.}
 \label{fig:allaws_dif}
\end{figure}

The empirical limits of the RC strip in the CMD could be defined by eye, as done in many works, but we have a more objective way to do it. We used Nishiyama's law with $\mu_{Ks}$\,=\,0.11 and 0.23\,mag\,kpc$^{-1}$ to encompass the RC overdensity - dashed lines in Fig.\,\ref{fig:RCstrip}. The fact that the RC strip crosses the Wd~1 MS, indicates that there are intruders in the overdensity strip. Many of those intruders can be excluded on the basis of their colours in the CCD. We used Nishyiama's law to evaluate the colour excesses and got the average $E_{J-Ks}/E_{H-Ks}$\,=\,2.9\,$\pm$\,0.07 for the RC candidates inside the strip. We used this average and excluded objects with measurements departing more than 3 $\sigma$, keeping the ones in the range 2.7\,$<$\,$E_{J-Ks}/E_{H-Ks}$\,$<$\,3.1, which resulted in 8463 stars with colours compatible with RCs, with $JHKs$ magnitudes. Of course, stars with similar colours but different absolute magnitudes (red dwarfs, K giants) remain in the RC sample. However, they work exactly in the same way as RC stars regarding the measurement of colour excess and the reddening law. 

For the RC candidates, we obtained $BVIZYW1W2$ magnitudes from the published catalogues described in Section\,\ref{section2.1}. The final number of RCs is: 26 in $B$, 64 in $V$, 205 in $I$, 1439 in $Z$, 1395 in $Y$, 8463 in $JHKs$ and 256 in $W1,W2$. We combine this RC sample with the 105 Wd~1 cluster members which have published spectral types and $BVRIJHKs$ magnitudes. Intrinsic colours for RC stars were taken from Table 1 in \citet{Nataf16} for the optical and NIR and from \citet{GF14} for WISE - their Table\,1 - interpolating the W1 and W2 effective wavelengths reported by \citet{Scha16} - their Table\,2.  Absolute magnitudes were calculated by adopting $M_{I,RC}$\,=\,$-0.12$ from \citet{Nataf13}. 
 
\subsection{Colour excess ratios combining Red Clump stars and Wd~1 cluster members}\label{section3.2}
The colour excess ratios $E_{J-H}\,/\,E_{J-Ks}$ are presented in Fig.\,\ref{fig:EJHXEJK} displays the colour excess ratio  for RCs (red crosses) and Wd~1 cluster members (blue crosses) and the linear fit relating these quantities. Similar fit for other colour excess ratios are presented in Appendix\,\ref{AppendixC}. We present below the set of colour excess ratios, combining data from RCs and the Wd~1 cluster members. 

\begin{eqnarray}
E_{B-J} &  = & (8.167\pm0.263) ~E_{J-Ks}, \label{eq3} \\
E_{V-J} &  = & (5.257\pm0.181) ~E_{J-Ks}, \label{eq4} \\
E_{R-J} &  = & (3.597\pm0.185) ~E_{J-Ks}, \label{eq5} \\
E_{I-J} & = & (2.465\pm0.032) ~E_{J-Ks}, \label{eq6} \\
E_{Z-J} & = & (1.797\pm0.012) ~E_{J-Ks}, \label{eq7} \\
E_{Y-J} & = & (0.841\pm0.004) ~E_{J-Ks}, \label{eq8} \\
E_{J-H} & = & (0.655\pm0.001) ~E_{J-Ks}, \label{eq9} \\
E_{H-Ks} & = & (0.345\pm 0.001) ~E_{J-Ks}, \label{eq10} \\
E_{J-H} & = & (1.891\pm0.001) ~E_{H-Ks}, \label{eq11} \\
E_{W1-J} & = & (-1.274\pm 0.030) ~E_{J-Ks}, \label{eq12} \\
E_{W2-J} & = & (-1.194\pm 0.055) ~E_{J-Ks}. \label{eq13} 
\end{eqnarray}

The set of colour excess ratios shows that our results are very accurate for the $JHKs$ bands, since we have much more data for these wavelengths than for shorter ones, and the relation for this set of three indices is completely dominated by RCs. The relation $E_{J-H} \times E_{J-Ks}$ is the tightest one. The plot even reveals the spread of extinctions among the Wd~1 cluster members and because of this we are anchoring our results on this particular colour ratio. The ratio $E_{J-H}/E_{J-Ks}$\,=\,0.655\,$\pm$\,0.001 corresponds to the more usual form $E_{J-H}/E_{H-Ks}$\,=\,1.891\,$\pm$\,0.001. 

In the case of $Z$ and $Y$ filters, we have only data for RCs (taken from the VISTA archive), not for the cluster members. Regarding the $R$ filter, on the other hand, we have data only for Wd~1. The corresponding relation $E_{R-J}$\,$\times$\,$E_{J-Ks}$ does not enable an accurate linear fit and we derived the angular coefficient from the average ratio between these colour excesses. The number of measurements from RCs in the $B$ and $V$ filters is much smaller than for longer wavelengths - because RCs are faint in that spectral range - and just the closer RCs have reliable photometry in those filters. As seen in the corresponding plots, the deep imaging of Wd~1 (reaching stars under large extinction)  was crucial to warranty a good fit in colour relations involving filters in the optical window.  

\subsection{Absolute extinctions derived from colour excess ratios}\label{section3.3}

\begin{figure}
 \centering
 \resizebox{\hsize}{!}{\includegraphics[width=20.0cm,angle=0, trim= 0cm 0.7cm 0cm 2cm, clip]{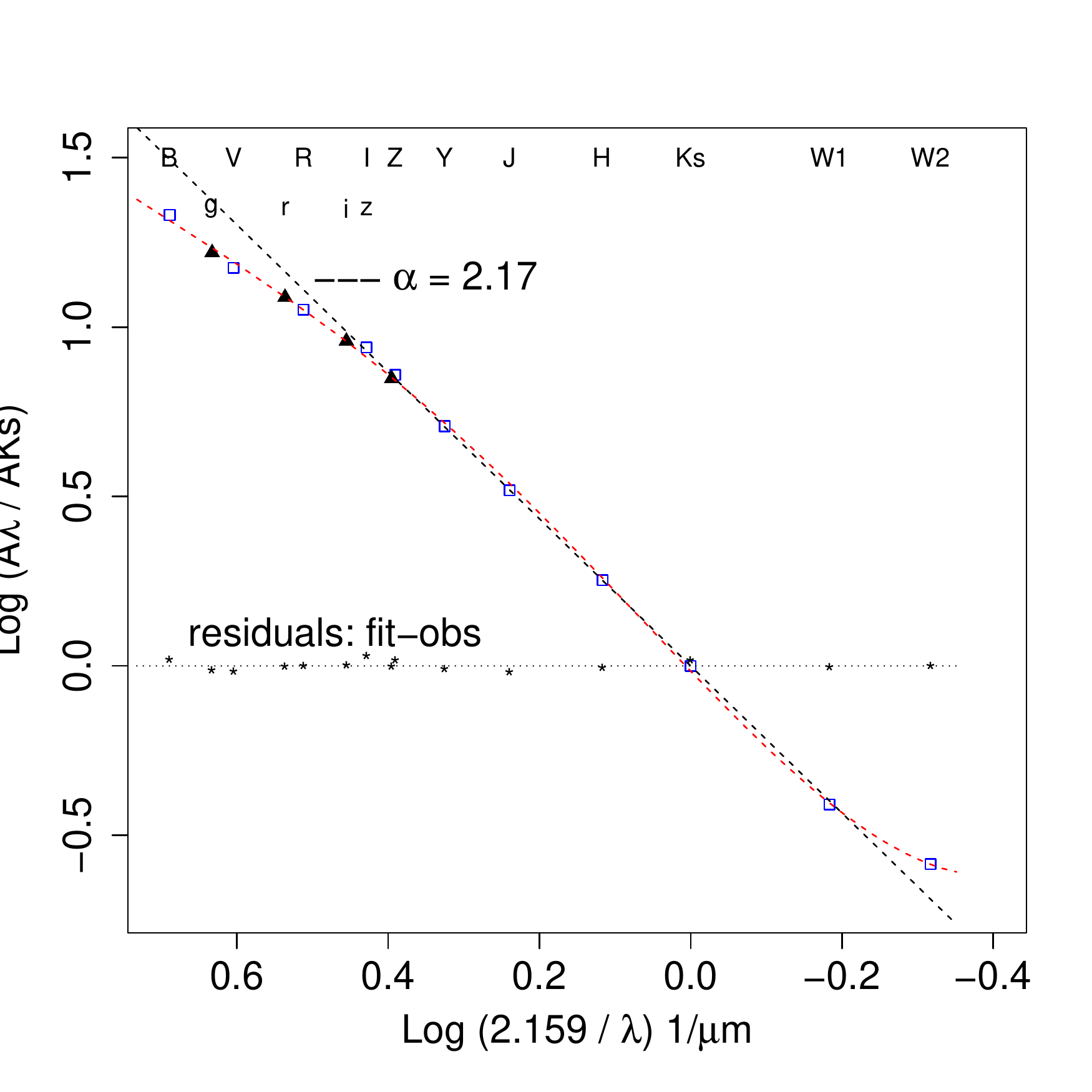}}
 \caption{Extinction law for the Galactic Plane (GP), combining our results (filters with capital letters at the top and empty squares) with colour excess ratios from \citet{Scha16} - filled triangles $g,r,i,z$ filters - for which we derived the extinctions. The dotted red line is the fit with Eq. \ref{eq19} and black asterisks are residuals from the fit minus observations. The black {\it dashed} line is the power law with exponent $\alpha$\,=\,2.17, which is a very good representation of the extinction law for the inner GP in the range 0.8-4.0\,$\mu$m.}
\label{fig:planelaw}
\end{figure}

We selected the colour excess ratio $E_{J-H}/E_{J-Ks}$ to derive the $\alpha$ exponent of the reddening law from Eq.\,\eqref{eq2}. This choice is based on the high accuracy of that index and on the fact that this is the best wavelength range to fit the reddening curve by a power law.
We performed an initial study to derive alpha based on the dynamical effective wavelengths, by following the procedure designed by Stead and Hoare (2009), which evolve as the colour indices increase. We approximate their relations from their Fig.\,2a by linear relations in Appendix~\ref{appendixA} to calculate the effective wavelengths as a function of $(H-K)$ for $JHKs$ filters. We then extracted the photometry from 2MASS catalogue inside a circle of 1$^\circ$ diameter centered on the Wd~1 cluster from which we selected RCs as we did for our photometry. We evolved the effective wavelengths using $H-Ks$ for each star, derived their $\alpha$ and took the median value: $\alpha$\,=\,2.25\,$\pm$\,0.15. 

We used an alternative and simpler procedure, as described in Section\,\ref{section2}, taking the wavelengths of each filter fixed, as derived by \citet{Stead+2009} using a K2\,III star. We found $\alpha$\,=\,2.14\,$\pm$\,0.10, in reasonable agreement with the more complex method and in excellent agreement with \citet{Stead+2009} who derived $\alpha$\,=\,2.03\,$\pm$\,0.18 from 2MASS data and $\alpha$\,=\,2.17\,$\pm$\,0.07 from UKIDSS data. Since our data are calibrated in the 2MASS system, we adopted the fixed effective wavelengths, as just mentioned, for our photometric data, which have an observational setup very close to that of 2MASS filters. Now, using Eq.\,\eqref{eq2} and the measured ratio $E_{J-H}/E_{J-Ks}$\,=\,0.655\,$\pm$\,0.001 we derived $\alpha$\,=\,2.126\,$\pm$\,0.080 which translates into $A_J/A_{Ks}$\,=\,3.229 and $A_H/A_{Ks}$\,=\,1.769.  

Using this value of $A_J/A_{Ks}$ and the colour excess ratios in the previous sub--section (Equations\,\ref{eq3} to \ref{eq13}), we derive the remaining $A_\lambda/A_{Ks}$ listed below:
\begin{eqnarray}
A_B = 21.43:A_V=14.95:A_R=11.25:A_I=8.72:  \nonumber \\
A_Z  = 7.23:A_Y=5.10:A_J=3.23:A_H=1.77:   \nonumber \\ 
A_{Ks}  =  1:A_{W1}= 0.39:A_{W2} = 0.26.
\label{eq14}
\end{eqnarray}

The above set of $A_\lambda/A_{Ks}$ defines an extinction law which is much steeper than typical ones published up to $\sim$10 years ago, but which are still in use \citep{CCM89,Indebetouw+2005,RL85}. From these relations, we can obtain the absolute extinction in all filters for a star, provided its $A_{Ks}$ is known. $A_{Ks}$ is easy to obtain from the colour excesses, for example: 
\begin{eqnarray}
A_{Ks} & = & 0.449 ~E_{J-Ks},	\label{eq15}	\\
A_{Ks} & = & 0.685 ~E_{J-H}, 	\label{eq16}	\\
A_{Ks} & = & 1.300 ~E_{H-Ks}.	\label{eq17}
\end{eqnarray}

In order to minimise the photometric errors, we use all these three relations and then perform the average to get a more robust value for $A_{Ks}$. We did this to obtain the extinction towards the Wd~1 cluster members, presented in Table\,\ref{table2} (column\,3) for all filters. The average value for this cluster based on 92 stars, after excluding those with obvious circumstellar contamination is:
\begin{equation}
<A_{Ks}~Wd1> = 0.736 \pm 0.079.
\label{eq18}
\end{equation}

The 18 WR stars which are not affected by dust emission resulted in $<A_{Ks} ~Wd1-WRs>$\,=\,0.736\,$\pm$\,0.082, which is indistinguishable from the less evolved  Wd~1 cluster members. This value is substantially lower than those derived by previous authors for this cluster, and this has an impact on the derived distance.

\subsection{A reddening law for the Galactic Plane in the range 0.4--4.8 $\mu$m}\label{section3.4}

Our target is located in the GP and it would be interesting to compare our results with others in the same region, since most of the recent studies based on large area surveys are focused on the Galactic Bulge (GB). A particularly important work in the GP is that reported by \citet{Scha16} for many thousands of stars, using the APOGEE spectroscopic survey and photometry from the ten--band Pan-STARRS1 survey.The majority of the targets are at -5$^\circ<$\,b\,$<$+5$^\circ$ and 0$^\circ<$\,l\,$<$250$^\circ$. \citet{Scha16} reported colour excess ratios in the optical, NIR and MIR windows, and their results are in excellent agreement with ours. They did not derive the extinction relations ($A_\lambda/A_{Ks}$), but this can be obtained in the same way as we described in the previous sub--section, especially because their ratios are anchored on 2MASS colours excesses. From their $E_{J-H}/E_{H-Ks} = 1.943$ we obtain $\alpha$\,=\,2.209, which translates into  $A_J/A_{Ks}$\,=\, 3.380 and $A_H/A_{Ks}$\,=\,1.809. The set of $A_\lambda/A_{Ks}$ corresponding to their colour excess ratios is: $\displaystyle A_g\,=\,16.61\,:\,A_r\,=\,12.24\,:\,A_i\,=\,9.10\,:\,A_z\,=\,7.05\,:\,A_J\,=\,3.38\,:\,A_H\,=\,1.81\,:\,A_{Ks}\,\,=\,\,1\,:\,A_{W1}\,=\,0.43\,:\,A_{W2}\,\,=\,\,0.21$. 
Due to the excellent agreement of their  results with ours, one can infer an average extinction law for the inner GP by performing a polynomial fit on both data sets in the range 0.4-4.8\,$\mu$m (see Eq.\,\eqref{eq19}, where  $x = log(2.159/\lambda)$ in units of $1/\mu$m, and the corresponding plot is in Fig.\,\ref{fig:planelaw}).

\begin{equation}
\log\frac{A_\lambda}{A_{Ks}} = -0.015 + 2.330 x + 0.522 x^2 - 3.001 x^3 + 2.034 x^4. 
\label{eq19}
\end{equation}

The standard deviation of the observed minus calculated fit (O--C) is very small and is driven by the residuals at wavelengths shorter than 1\,$\mu$m, specially in the $B$ filter. It is surprising to see that a power law with $\alpha$\,=\,2.17 is an almost perfect representation of the data in the range 0.8-4\,$\mu$m with r.m.s.\,=\,0.09\,magnitudes. 

\section{Comparison between different extinction laws}\label{section4}
\subsection{The $\alpha$ exponent and JHKs colour ratios}\label{section4.1}

The comparison between reddening laws from different authors should be straightforward in the NIR, since $JHKs$ filters are very similar in modern times. Most of the authors report a narrow range of $E_{J-H}/E_{H-Ks}$ values (1.8--2.0). Surprisingly, there is a large spread in the published reddening laws, even when derived in the narrow NIR window and also when using the same database (2MASS): $\alpha$\,$\approx$\,1.65-2.64, which translates into $A_J/A_{Ks}$\,$\approx$\,2.5-4.2. The different procedures used to derive the extinction law from the colour excess ratios have different biases. In most cases, this is due to the difficulty in transforming broad-band measurements into monochromatic wavelengths \citep{Sale15}, but there are other additional issues, some of which we describe here. 

\begin{figure}
 \centering
 \resizebox{\hsize}{!}{\includegraphics[width=20.0cm,angle=0, trim= 0cm 0.7cm 0cm 2cm, clip]{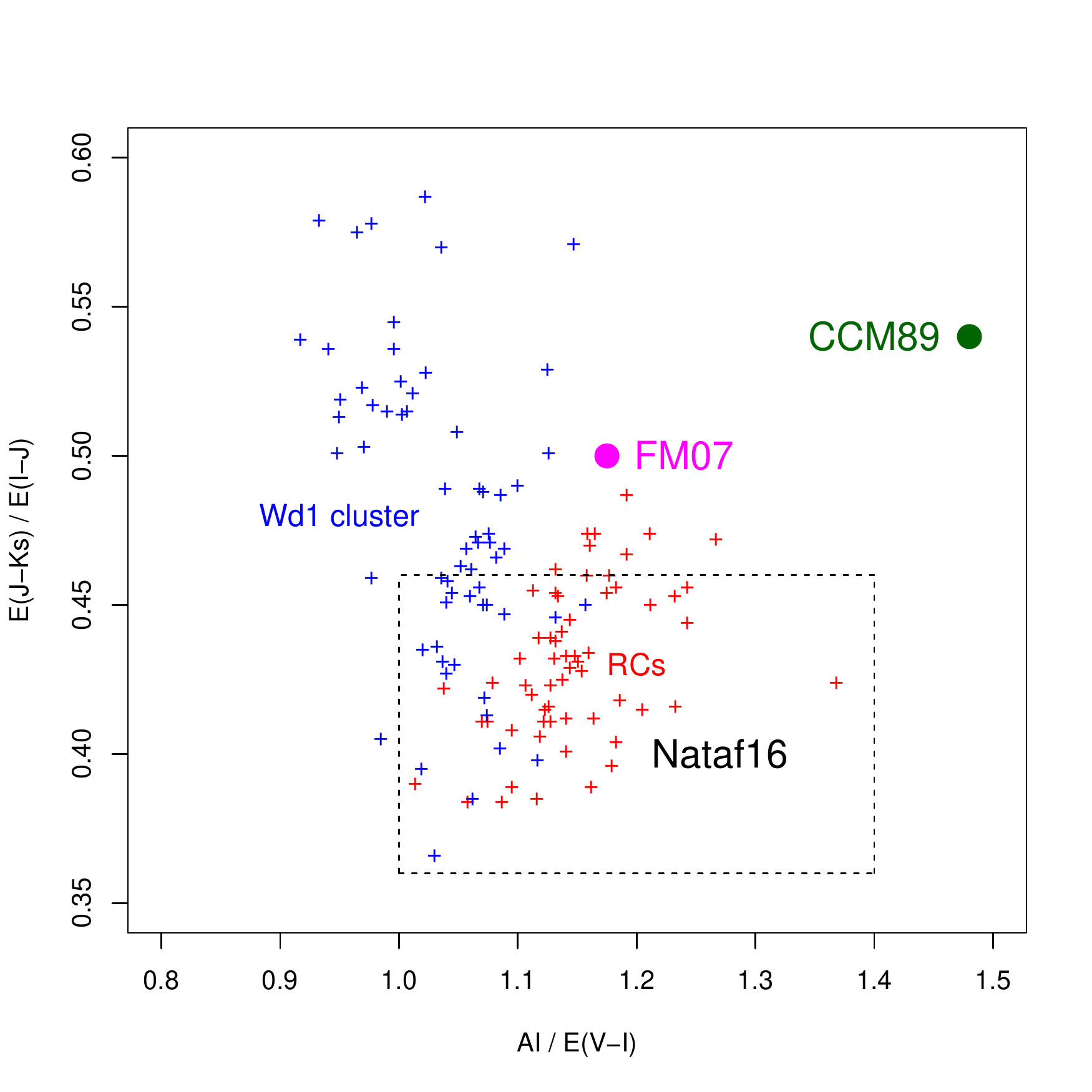}}
 \caption{Comparison between our results with other authors, showing a lack of correlation between $E_{J-Ks}/E_{I-J}$ and $A_I/E_{V-I}$. Our result for RCs (red crosses) compares well with that by \citet{Nataf16} - dashed rectangle - for the Galactic Bulge and samples the low density, higher ionization ISM.  Wd~1 cluster members (blue crosses) have a contribution from a slightly different grain distribution from a denser medium. Older results by \citet{FM07} and by \citet{CCM89} are represented by pink and green circles, respectively.}
 \label{fig:RJKIJxRAIVI}
\end{figure}

\begin{table*}
\scriptsize
\label{table1}
\begin{minipage}{170mm}
\caption{Comparison between some reddening laws. Authors - 1st row: vdH\#16 = \citet{Hulst1946}, Nat16 = \citet{Nataf16}, Sch16 = \citet{Scha16}, SH09 = \citet{Stead+2009}, Ind05 = \citet{Indebetouw+2005}, CCM89 = \citet{CCM89}, FM09 = \citet{FM09}, Nishi06 = \citet{Nishiyama+2006}, Nishi08 = \citet{Nishiyama+2008}.
 Values marked with * were derived in this work from reddening and effective wavelengths reported by the corresponding authors. The second row indicates the photometry source. For FM09, we used only the star BD+45 973.}
\begin{tabular}{lcccccccccccc}
\hline
\hline
Filter               &$\lambda_{\rm eff}$& Wd~1         & Wd1  + RCs & vdH\#16&Nat16&Sch16& SH09& Ind05& CCM89& FM09& Nishi06,08 \\
\hline
                  &  $\mu$m  &$A_{\lambda}$&$A_{\lambda}/A_{Ks}$&&VVV  & Pan-STARRS1&UKIDSS &2MASS   &&HST/SNAP & SIRIUS \\
\hline
$B$                             &  0.442 &15.49$\pm$0.10&       21.43& 14.36 &  -   & -      &-    &  -  & 11.49& 17.52& -\\
$V$                             &  0.537 &11.26$\pm$0.07&       14.95&  11.69& 13-15& 16.61g*&-    &  -  &  8.77&12.76 &16.13\\
$R$                             &  0.664 &8.44$\pm$0.10 &       11.25&   8.63& -    & 12.24r*&-    & -   &  6.59&  9.48& -\\
$I$                             &  0.805 &5.71$\pm$0.04 &        8.72&   6.32& 7.26 & 9.10i* &     & -   &  -   &  6.52& -\\
$Z$                             &  0.878 &  -           &        7.23&   5.30&  -   & 7.05z* & -   &  -  &  4.20&  5.81& -\\
$Y$                             &  1.021 &  -           &        5.10&   3.97&  -   &  -     &  -  &-    &  -   &  -   & -  \\
$J$                             &  1.244 &2.34$\pm$0.03 &        3.23&   2.70& 2.85 &  3.56* &3.25*&  2.5& 2.47&  2.90& 3.02 \\
$H$                             &  1.651 &1.29$\pm$0.02 &        1.77&   -   &  -    &  1.84*&1.78*&1.61 & 1.54 &  1.67&1.73 \\
$Ks$                            & 2.159  &0.74$\pm$0.01 &       1    &   1   &  1    &   1   & 1   &    1&    1 &  1   & 1  \\
$W1$                            & 3.295  &0.29$\pm$0.03 &        0.39&  -     &  -   &0.43*  & -   &  -  &    - &  -   &0.40 \\
$W2$                            & 4.4809 & 0.19$\pm$0.05&        0.26&   -    &  -   & 0.21* & -   & -   &    - &  -   &0.20 \\
\hline
{\bf$\alpha$}                   &  -     &  -           &        2.13&   -    &1.88* & 2.21*& 2.14  &1.66*&1.68*& 2.00& 1.99 \\
\hline
\end{tabular}

\end{minipage}
\end{table*}

An example indicating differences related to the methodology is by comparing three extreme results based on 2MASS data. We adopted the effective wavelengths based on \citet{Stead+2009}, reported in Section\,\ref{section2}, and the observed colour excess ratio $E_{H-Ks}/E_{J-Ks}$\,=\,0.345 to derive $\alpha$\,=\,2.13 and $A_J/A_{Ks}$\,=\,3.23. \citet{Indebetouw+2005} measured $E_{H-Ks}/E_{J-Ks}$\,=\,0.36 and derived $A_J/A_{Ks}$\,=\,2.5 directly by minimization of the colour excesses. It is surprising to see that \citet{GF14} using 2MASS data obtained a very high $\alpha$\,=\,2.64 ($A_J/A_{Ks}$\,=\,4.2), although their colour excess ratio $E_{J-H}/E_{H-Ks}$\,=\,1.934 is very similar to others. Their method is complex to follow what exactly drove that power law exponent, but it corresponds to the the extreme values reported by \citet{FM09}, from which they adopted the reddening law. In some way their procedure converged to the higher and not to the average exponent of \citet{FM09}. Using our procedure (anchoring the reddening law in the $\alpha$ exponent) with the colour excess ratios reported by those authors, we derive $\alpha$\,=\,1.89 with $A_J/A_{Ks}$\,=\,2.84 and $\alpha$\,=\,2.15 with $ 3.27$, for \citet{Indebetouw+2005} and \citet{GF14} respectively. This agreements indicates that small differences on the effective wavelengths are not the main cause of the discrepancies found in the reddening laws.

Beyond discrepant methodolgies, genuine differences due to dust properties do exist and are clearly shown in \citet{FM09}, based on spectrophotometric studies and not affected by problems defining the effective wavelengths. Large area surveys based on broad--band filters, such as reported by \citet{Nataf16} for the Galactic Bulge and by \citet{Scha16} for the GP, also show real variations and surprisingly, they exist on very small scales (arcminutes), on top of a general pattern related to the angular distances from the Galactic Centre. 

\citet{Nishiyama+2006} used their SIRIUS survey of the Bulge to measure  $E_{H-Ks}/E_{J-Ks}$\,=\,0.343 and derived $\alpha=1.99$ with $A_J/A_{Ks}$\,=\,3.02, in reasonable agreement with our results for the Wd~1 direction. Our extinction law is in excellent agreement with \citet{Stead+2009}, who obtained $\alpha$\,=\,2.14.
By translating their exponential law into extinction ratios (not presented in their original work), we computed $A_J/A_{Ks}$\,=\,3.25.
\citet{Nataf16} reported an $<A_J/A_{Ks}>$\,=\,2.85 from the large and deep VVV survey towards the Bulge, as in \citet{Nishiyama+2006}, but covering more diverse environments close to the GP. Using the effective wavelengths from VVV, this corresponds to $\alpha$\,=\,1.88.
\citet{Scha16} measured $E_{J-H}/E_{H-Ks}$\,=\,1.943 for a 2MASS sample in the inner GP, which corresponds to $E_{H-Ks}/E_{J-Ks}$\,=\,0.340. Using the wavelengths they adopted for 2MASS effective wavelengths ($\lambda_J$\,=\,1.2377, $\lambda_H$\,=\,1.6382 and $\lambda_{Ks}$\,=\,2.151) this implies $\alpha$\,=\,2.3 with $A_J/A_{Ks}$\,=\,3.56 and $A_H/A_{Ks}$\,=\,1.84. This is a little higher than our value, and using our adopted effective wavelengths this is reduced to $\alpha$\,=\,2.21 with $A_J/A_{Ks}$\,=\,3.38 plus $A_H/A_{Ks}$\,=\,1.81. We will use these last values in Fig.\,\ref{fig:allaws_dif} for consistency with our procedure, although differences are very small.
As a summary we see that in general, extinction laws in $JHKs$ bands derived after 2005 are steeper than previous ones and this is driven by the deep surveys in the inner Galaxy.

The foreground extinction we obtained in this work for the Wd~1 cluster, $A_{Ks}$\,=\,0.736, is smaller than $\approx$\,0.9-1.0 claimed in previous papers \citep{Andersen16, Gennaro+2011, Crowther+2006, Brandner+2008}. We derived a separate $A_{Ks}$ extinction for 18 Wolf--Rayet stars (excluding the dusty WC9) and obtained $A_{Ks}$\,=\,0.743\,$\pm$\,0.08 in excellent agreement with that based on non--WR cluster members. For comparison, using the same set of WRs, \citet{Crowther+2006} reported $A_{Ks}$\,=\,0.96\,$\pm$\,0.14. Later, those authors revised their value to $A_{Ks}$\,=\,1.01\,$\pm$\,0.14 \citep{Crowther+2008}. A recent study on the Wd~1 low mass contents by \citet{Andersen16} reports $A_{Ks}$\,=\,0.87\,$\pm$\,0.01 in better agreement with ours than previous works, but still not compatible. This is because they used the freddening line from \citet{Nishiyama+2006}, which is appropriate for the Bulge and our target in in the GP. \citet{Andersen16} found some evidence for the extinction gradient to increase towards N--NE, based on a large population of PMS, in agreement with our result, base on fewer evolved stars. Those authors decided to derive the extinction for every single PMS star, instead of using an average extinction.

Our absolute extinction for Wd~1 is not compatible with previous ones and has an impact on the cluster distance, age and the absolute magnitudes of WRs. After measuring the cluster distance based on an eclipsing binary, we will tackle this question in a future paper.

\begin{figure}
\centering
 \resizebox{\hsize}{!}{\includegraphics[width=20.0cm,angle=0]{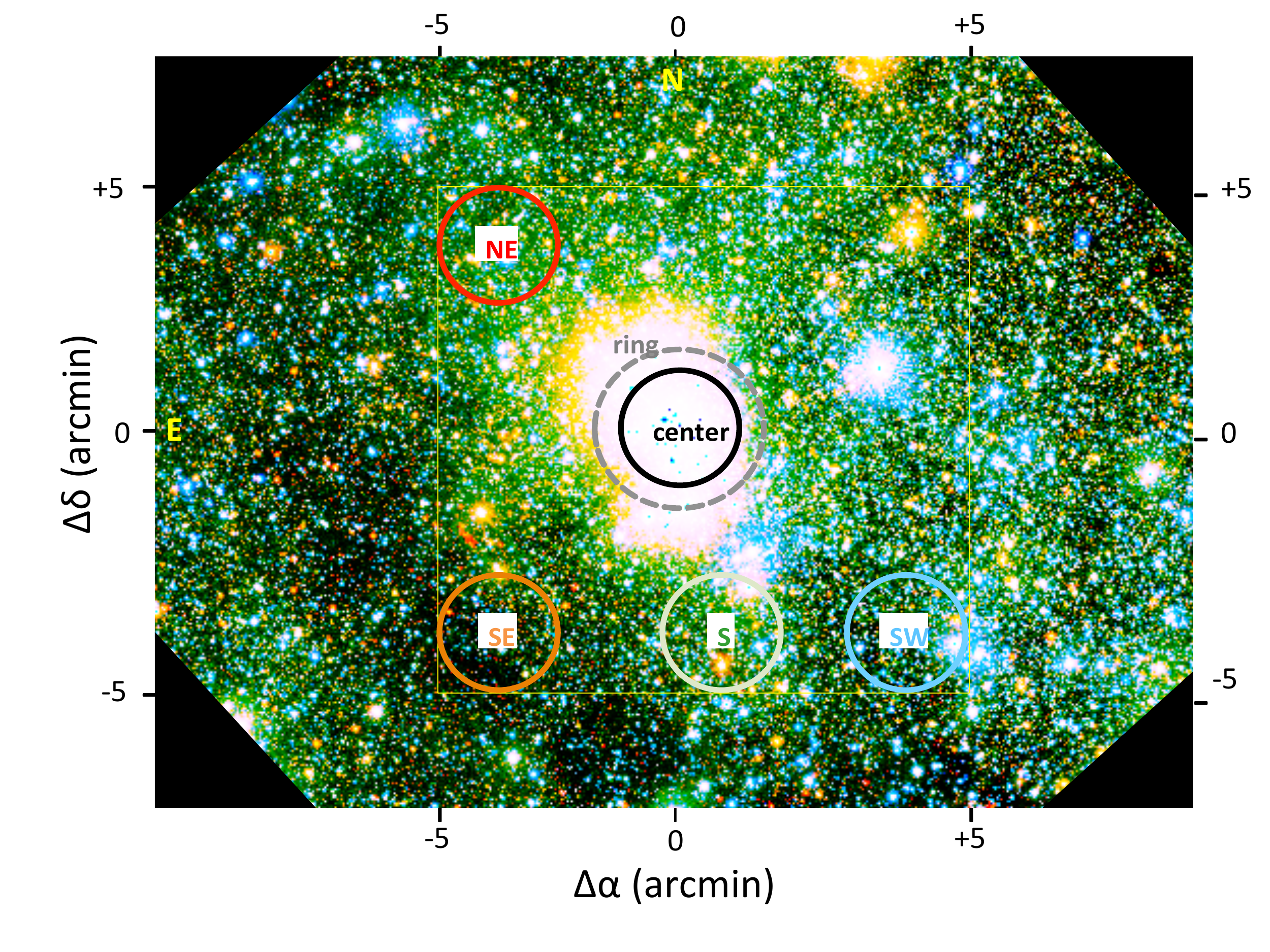}}
 \caption{Colour coded map from VVV survey from filter $Z$ (blue), $J$ (green) and $Ks$ (red). The labels indicate the directions in Fig.\,8 for which we measured the density of stars as a function of $J-Ks$ colour. The circles have $r$\,=\,2\arcmin. The {\it Centre} contains the Wd~1 cluster and the {\it ring} has width\,=\,1\arcmin. The yellow rectangle indicates the ISPI FOV.}
 \label{fig:VVV_map}
\end{figure}

\begin{figure}
 \centering
 \resizebox{\hsize}{!}{\includegraphics[width=20.0cm,angle=0, trim= 2.2cm 0cm 3.6cm 0cm, clip]{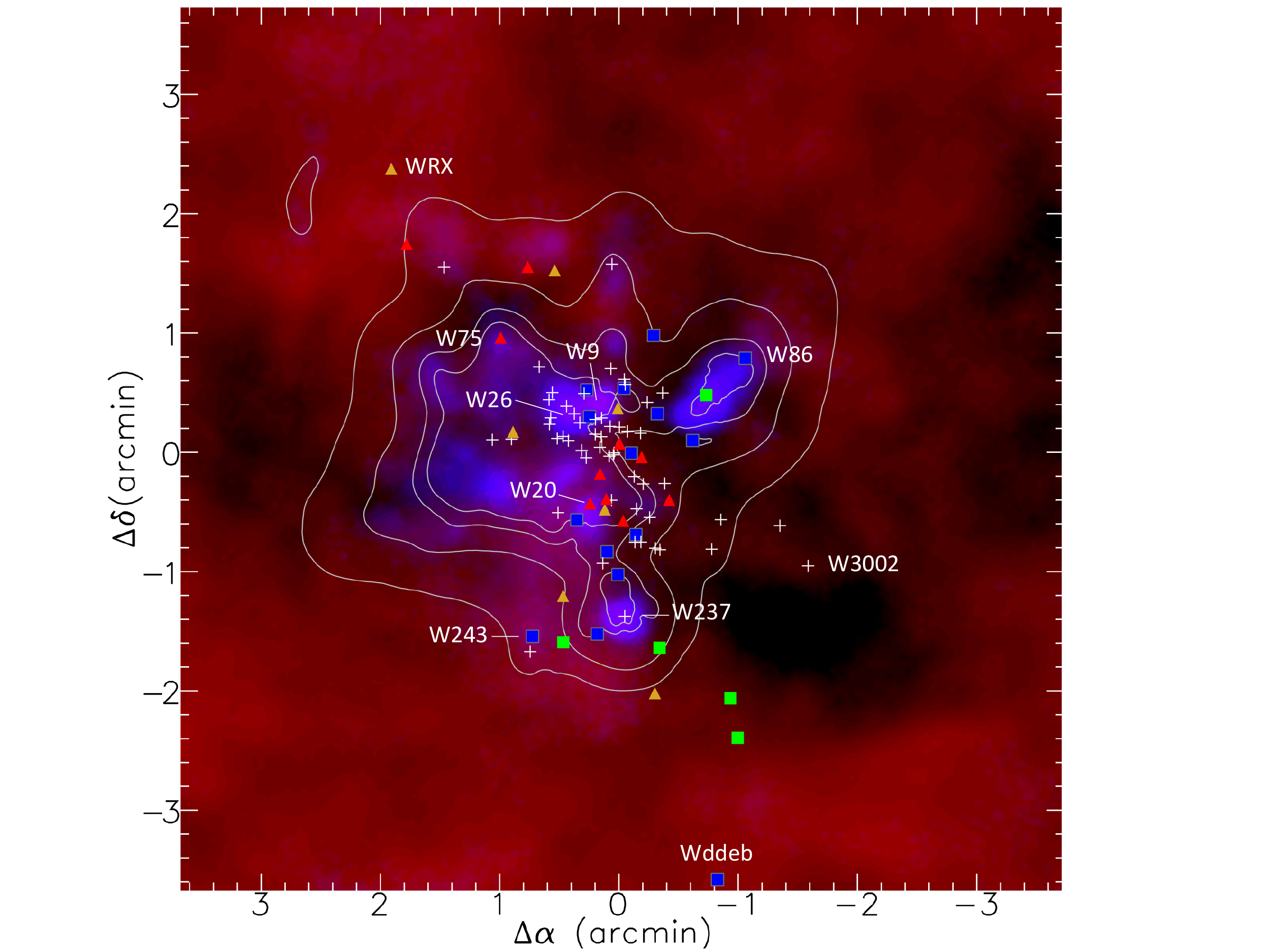}}

 \caption{  Extinction map. a):  $A_{Ks}$ for the cluster members: {\it white crosses} are for values departing less than 0.5$\sigma$ from the average ($0.70<A_{Ks}<0.77$); {\it orange triangles} for slightly  higher reddening ($0.78<A_{Ks}<0.79$); {\it red triangles}   for the highest valus ($0.81<A_{Ks}<1.17$); {\it green squares} for $0.66<A_{Ks}<0.70$; and {\it blue squares} for  $0.60<A_{Ks}<0.65$. b): {\it red clouds} for cold dust (160~$\mu$m); {\it blue clouds} for warm dust (70~$\mu$m) from Herschel survey; and {\it contour lines} for hot dust (24~$\mu$m) from WISE survey. Black regions in the Figure indicate the absence of dust emission. For a complete identification of the features in this Figure, we refer to the papers \citet{Negueruela+2010}, \citet{Ritchie+2009} and \citep{dougherty10}.
 }
\label{fig:dustmap}
\end{figure}

\subsection{Relation between the extinction in the NIR, optical and MIR windows}\label{section4.2}

As seen in Fig.\,\ref{fig:allaws_dif} all recent laws are much steeper than that of \citet{Hulst1946}, taken as a zero point, in contrast with \citet{CCM89}, which is much shallower. The frequently used rule of thumb $A_V/A_{Ks}$\,$\approx$\,10 based on that law and other similar laws, is no longer acceptable as representative for the Galaxy . The derived ratio in this work is $A_V/A_{Ks}$\,=\,14.95 and that of \citet{Scha16}, at a little shorter wavelength is $A_g/A_{Ks}$\,=\,16.61, both done along the GP. It is surprising that these values are not far from $A_V/A_{Ks}$\,$\approx$\,13-16 we derived from the colour excess ratios reported by \citet{Scha16} and  $A_V/A_{Ks}$\,=\,16.13 reported by \citet {Nishiyama+2008}, both for the Bulge.

The absolute extinction towards Wd~1 obtained in the present work $A_V$\,=\,11.26\,$\pm$\,0.07 is in excellent agreement with $A_V$\,=\,11.6 derived by \citet{Clark+2005} for Yellow Hypergiants in Wd~1.

After the works by \citet{FM09} and \citet{Nataf16}, it became clear that all families of extinction laws cannot be represented by a single parameter, $R_V$, for example. \citet{Nataf16} showed that the the relation the ratios $R_{JKIJ}$\,=\,$E_{J-Ks}/E_{I-J}$ and $R_I$\,=\,$A_I/E(V-I)$ are not correlated. This is not unexpected, since the grain properties might be different in the general field (RCs) and inside the denser environment of the cluster. 
In Fig.\,\ref{fig:RJKIJxRAIVI} we compare our results with those from previous works. 

Wd~1 cluster members are seen under a range of extinction $A_V$\,$\approx$\,9-15 magnitudes, indicating the presence gas/dust condensations in the intra-cluster medium.
Results based on alternative ratios in the optical (e.g. $A_V/E_{B-V}$ and $A_V/E_{V-I}$  against $R_{JKIJ}$) point to the same scenario of a lower $A_V/E_{B-V}$ in the inner Galaxy as shown by \citet{Nataf13}.The $R_V$\,=\,$A_V/E_{B-V}$ ratio have been reported to disagree with the ``universal'' $R_V$\,=\,3.1 for the Wd~1 cluster \citep{Clark+2005,Negueruela+2010,Lim+2013}. Using 38 members of the Wd~1 cluster we obtained $\displaystyle R_V = {A_V} / {E_{B-V}} = 2.50\pm0.04$. 

For the MIR, using data from WISE for our measured colour excess ratios, we obtained $A_{W1}$\,=\,0.39\,$\pm$\,0.03 and $A_{W2}$\,=\,0.26\,$\pm$\,0.02. 
Such values are in very good agreement with those derived from \citet{Scha16} colour excess ratios.

\section{Extinction map}\label{section5}

The extinction map towards the Wd~1 cluster can be tackled in several diffrent ways: a) by the distrubution of the cluster members extinction; b) from maps of the colours of stars; and c) from the correlation between these two maps and the distribution of dust clouds in the FOV.  

\subsection{Extiction map of the cluster members}\label{section5.1}

In Fig.\,\ref{fig:dustmap} we present a map of extinction towards Wd~1. Points represent our measurements of $A_{Ks}$ for cluster members. White crosses represent the average extinction ($0.66<A_{Ks}<0.82$) or $10.1<A_{Ks}<12.6$), orange and red triangles higher than 1$\sigma$ and green and blue points are for values  lower than 1$\sigma$ from the average. Lower extinction dominates at W/SW regions and higher extinction to N/NE from the cluster centre and in the central region there are all ranges of values.  The mixture of values in the central region indicates the existence of intra--cluster dust, with some members displaced to our side and others to the back. It seems that at W/SW the dust is in the back side of the cluster, but crossing to the front at N/NE. The reality of this gradient can be checked in the next sub--section, when we discuss the far--infrared (FIR) images. 
Typical cluster members are seen through extinction in the range $0.66<A_{Ks}<0.82$ or $10.1<A_V<12.6$, represented by the  1$\sigma$  spread around the median value (78 stars). They are affected by interstellar plus local extinction. The $A_V = 2.5$ magnitudes range is due to local dust in front or internal to the cluster, and is surpringly large for $\sim 3.5 \arcmin$ FOV. The interstelar component can be measured from the (five) stars with lower extinction: $A_{Ks}=0.63\pm 0.02$ or $A_V=9.7\pm 0.30$. The nature of the reddening for the highest  (seven) values $0.83<A_{Ks}<1.17$ or $12.7<A_V<17.9$ is difficult to assess: some stars may have really high extinction, but others might be affected by hot circumstellar dust emission, which mimics large reddening.
\subsection{Cold and warm dust probed by FIR imaging }\label{section5.2}
The red clouds in Fig.\,\ref{fig:dustmap} represent cold dust as measured by Herschel survey as measured by Herschel survey, using the PACS instrument \citep{Poglitsch2010}.
It is notable that there are areas clear of dust (black regions) at E and W, probably caused by winds from massive stars and/or supernovae explosions. Just three "fingers" of cold dust reach the central regions of the cluster at NW, S and N and their tips are crowned by bright blue spots, tracing arm dust. The ''elephant trunk" at NW  was detected in radio emission by \citet{dougherty10}.  Warm dust impacts the cluster center, but dominates the East side, in rough agreement with higher extinction measured for stars. Four noticeable spots of warm dust coincide with the Red Supergiants (RSGs) W237, W20, W20 and the W9 B[e] star. However, the reddening of the RSGs stars is not larger than the average, indicating that the heated dust there is in the back of the star cluster. 
 
\subsection{Extinction and colours in the field around Wd~1}\label{section5.3}

We can use the spatial density of the colour indices $(J-Ks)$ in our whole catalogue of sources to study the 3D distributon of dust. 
To study the surface density of stars projected inside a particular FOV, we need to keep in mind the many effects involved in the observed star counts. The most important parameter is the increase in the projected density with the distance, which grows as the volume of the spherical sector in the FOV (for uniform density). At a given magnitude limit this is counterbalanced by the inverse square law for the brightness and the extinction. Finally, in some regions and at a given image quality, stellar crowding drops the stellar counts. These considerations can provide guidelines to interpret the stellar counts as a function of magnitude in a particular image, but most of the effects can not be corrected. 

\begin{figure}
\centering
\resizebox{\hsize}{!}{\includegraphics[width=20.0cm,angle=0, trim= 2.6cm 0.3cm 2.6cm 0.5cm, clip]{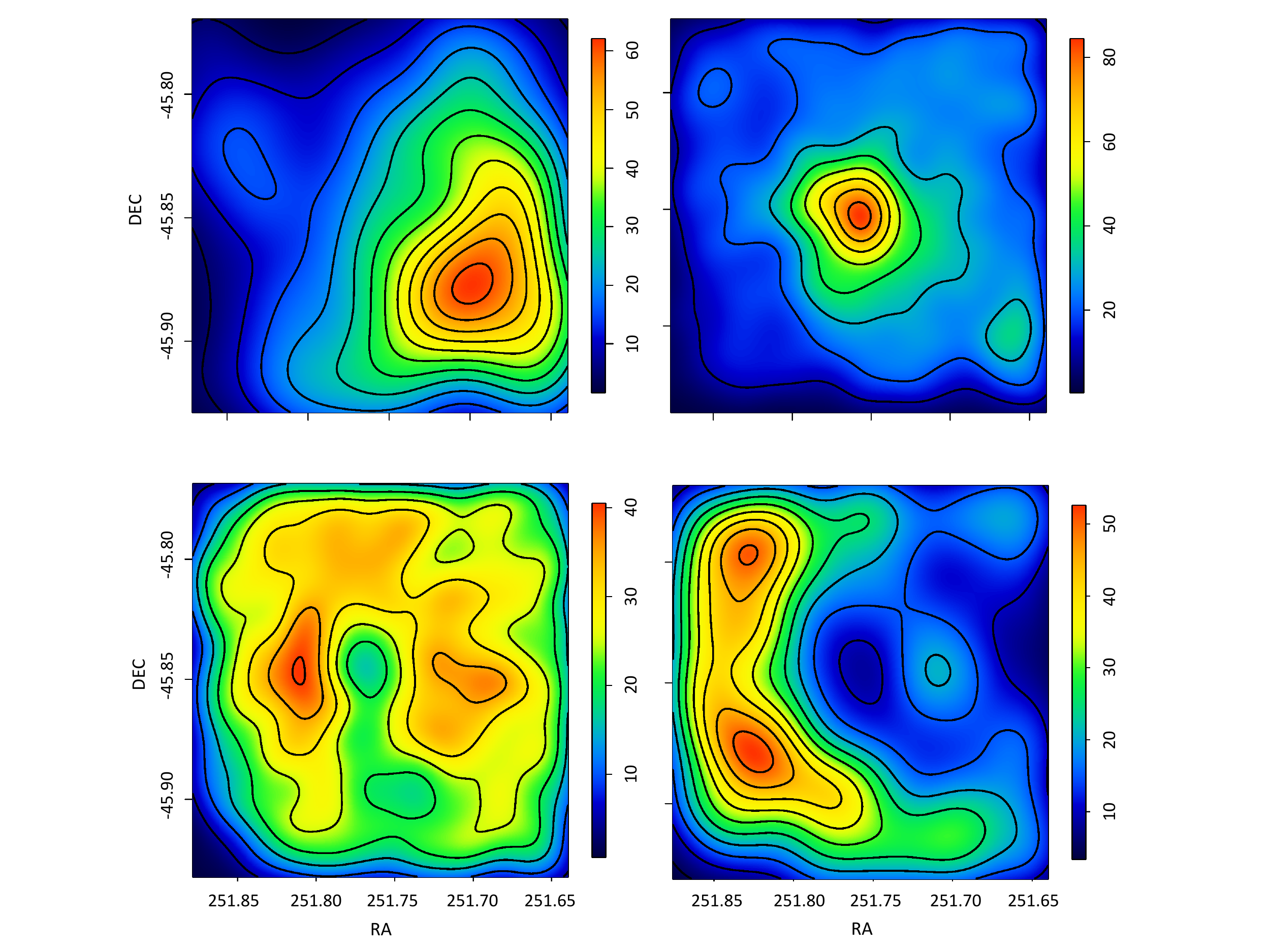}}
\caption{Stellar density maps:redder zones represents higher stellar densities. {\it Top left}:  0.350\,$>$\,$J-Ks$\,$>$\,0 showing a foreground concentration of stars with low extinction at SW of Wd~1 cluster. {\it Top right}:  1.788\,$>$\,$J-Ks$\,$>$\,1.251 showing the Wd~1 cluster with an extended halo. {\it Lower left}: 3.041\,$>$\,$J-Ks$\,$>$\,2.145 located farther than Wd~1. The low density region coincident with the cluster position is due to photometric incompleteness caused by crowding. The similar one to the S of the cluster is due to a cloud of higher extinction. {\it Lower right}: 4.365\,$>$\,$J-Ks$\,$>$\,3.757: the higher densities at the E and SE of the cluster are due to a zone of enhanced extinction farther than Wd~1.}
\label{sb1:colour_density}
\end{figure}

We binned the $J-Ks$ colour indices for ISPI+CAMIV photometry and normalised each slice of colours by the average colour index of the entire slice. After some initial trials, we selected a few slices which are representative of the features we found in the image, as shown in Fig.\,\ref{sb1:colour_density}.The upper left panel of in Fig. \ref{sb1:colour_density} displays a remarkable concentration of blue stars at W--SW of the Wd~1 which is probably due to a real cluster and not a result of a window of low extinction on the image.
The  upper right panel (1.251\,$<$\,$(J-Ks)$\,$<$\,1.788) shows the Wd~1 cluster at the centre of the image. The cluster size is larger than 6\arcmin and is elongated in the way described by \citet{Gennaro+2011} for the inner regions. The lower left panel shows a slice without strong concentrations in the stellar density. The regions of enhanced density at E and W of the cluster centre seems to be a halo of red pre main sequence (PMS) stars belonging to the cluster. The lower density at the S is probably due to a high extinction patch which shifts stars from that particular slice of colours to higher values since on average that slice represents the stellar field farthest  from the Wd~1 cluster ($D$\,$>$\,4.5\,kpc). The depression at the Centre is due to photometric incompleteness caused by crowding of cluster members. 

The lower right panel (3.757\,$<$\,$(J-Ks)$\,$<$\,4.365) represents a slice of very red colour indices. In addition to distant stars (reddened by distance), it also represents stars shifted to redder colours by clouds both closer and farther than Wd~1. This is the case for the enhanced density at S of Wd~1. We suggest that the enhanced density zone ranging from SE to NE is also due to an obscuring cloud more distant than the central cluster.

\begin{figure}
 \centering
 \resizebox{\hsize}{!}{\includegraphics[width=20.0cm,angle=0, trim= 0cm 0.5cm 0cm 2cm, clip]{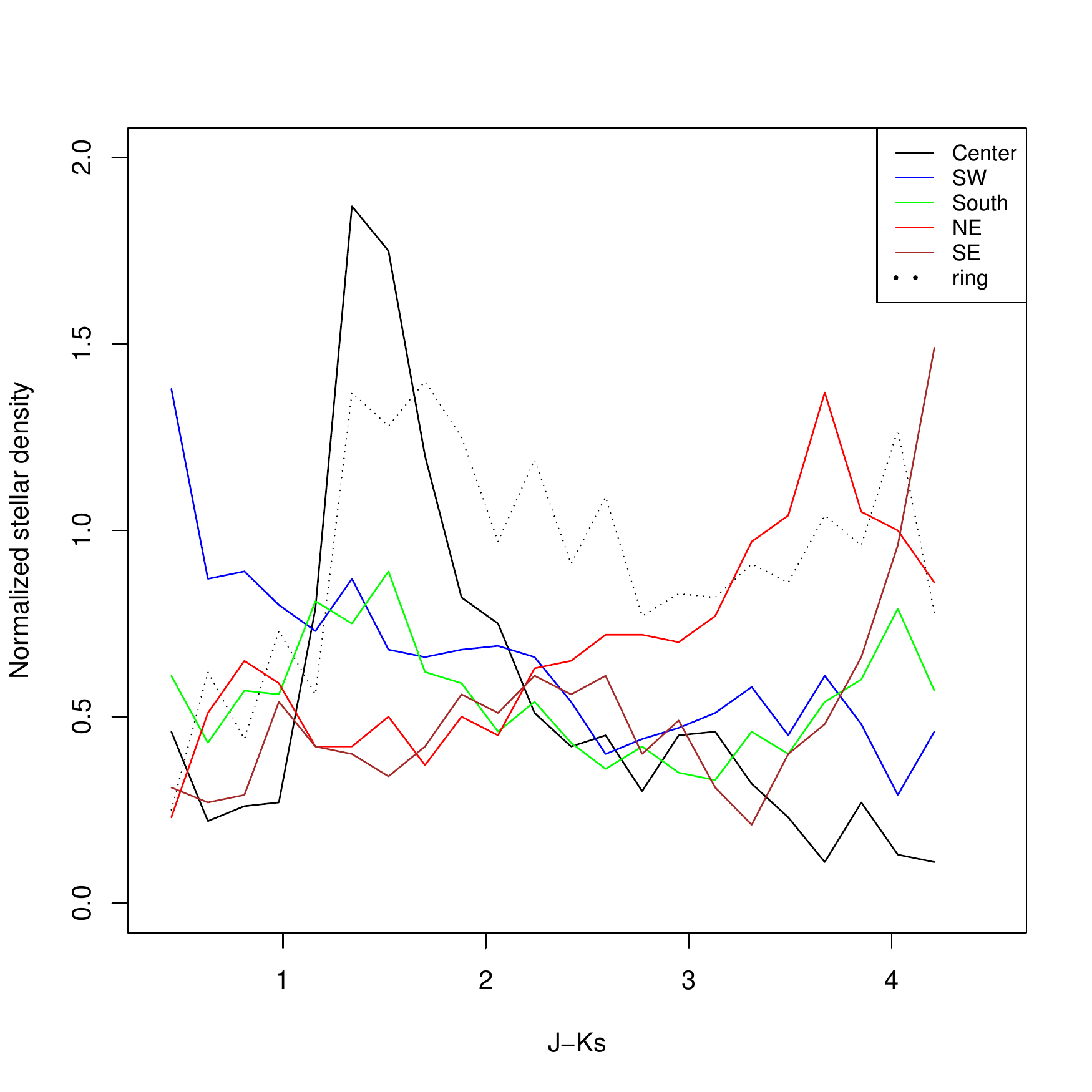}}
 \caption{Density of stars for a range of bins in $J-Ks$ colour in some key directions defined in Fig.\,\ref{fig:VVV_map}. The first maximum in the {\it blue} line indicates spatial overdensity of stars at the SW foreground. The peak of the black {\it solid} and {\it dotted} lines ($J-Ks$\,$\approx$\,1.6) indicates the cluster centre. The green line (S) shows an increasing density towards redder colours up to the cluster colour/distance, then a decline followed by a recovery for large colour indices. For the NE and SE directions (red and brown lines) there is a similar behaviour, but the enhanced extinction occurs at the cluster distance/colour.}
\label{fig:J-Kdensity}
\end{figure}

In order to explore in more detail these results, we made a plot with VVV images of the region, as seen in Fig.\,\ref{fig:VVV_map}. In fact there is a zone with blue stars (foreground young stars) at W-SW and a dark patch at the S and E of the cluster, indicating zones of higher extinction. The other structures revealed in Fig.\,\ref{fig:J-Kdensity} are not seen in the colour image of Fig.\,\ref{fig:VVV_map}. We then selected a few spatial directions, shown in the VVV map (Fig.\,\ref{fig:VVV_map}), to measure the stellar density as a function of the colour index $J-Ks$ along the cylinders with 2$'$ circular sections. In Fig.\,\ref{fig:J-Kdensity} we plot the density profiles with the same colours as the circles in Fig.\,\ref{fig:VVV_map}.
The solid black line in Fig.\,\ref{fig:J-Kdensity} shows the density profile in which the most remarkable feature is the cluster in the colour range 1\,$<$\,$J-Ks$\,$<$\,2. The red extension of the peak is due to the PMS stars, which are intrinsically redder than the MS members in the peak. PMS stars also show up in the ring around the cluster area ({\it black dotted line}), with colour up to $J-Ks$\,$\approx$\,2.5. The {\it blue line} in both Figs.\,\ref{fig:VVV_map} and \ref{fig:J-Kdensity} shows the concentration of stars with $J-Ks$\,$<$\,1 -- bluer than the rest of the sources -- indicating that they are foreground stars, as discussed above. The {\it green line} explores the density profile to the S of the cluster. The density profile in this direction is normal for colours bluer than those of the Wd~1 cluster, but than it decreases faster than  normal up to $J-Ks$\,$\approx$\,3 when the stellar density grows again up to  $J-Ks$\,$\approx$\,4. The minimum around $J-Ks$\,$\approx$\,3 is caused by enhanced extinction, which shifts the counts to larger reddening. The {\it red line} is for the circle at NE of  the cluster. Its behaviour is normal for $J-Ks<1$ after which there is a local minimum in coincidence with the cluster colour, followed by a constant increase up to $J-Ks$\,$\approx$\,3.8. Our interpretation is that there is an enhanced extinction at the distance of the cluster in that direction. The {\it brown line} corresponds to the circle at the SE and behaves similarly to that at NE, except it starts with lower counts than at the NE and presents two minima around $J-Ks$\,$\approx$\,1.5 and 3.5. Our interpretation is that there are two clouds at different distances in this direction. The extinction toward the SE looks to be higher than at any other direction. Just for a sake of spatial scale, the colour excess grows in proportion to  $E_{J-Ks}$\,$\approx$\,0.37\,mag\,kpc$^{-1}$, in manner that the deepest minimum at $J-Ks$\,$\approx$\,3.7 corresponds to a distance {$D$\,$\approx$\,10\,kpc.
 
As a summary, the most obscuring clouds are located to the S--SE border of our image at $D$\,$\approx$\,10\,kpc and there is a lighter obscuring cloud at SE--NE border at a distance comparable to that of Wd~1, but which does not impact the borders of the ISPI image at S--SE. The question is if this particular cloud has any impact on the properties of the stellar cluster.

\section{The 8620\,\AA\,DIB tracing the extinction beyond $E_{B-V} > 1$}\label{section6}

 \citet{Munari+2008} and \citet[][]{Wallerstein+2007} showed that the 8620\,\AA\,DIB has a tight correlation with $E_{B-V}$. \citet{Maiz+2015} also showed the linear correlation of the equivalent width of this DIB (EW8620\,\AA) as a function of the extinction holds up to $A_V$\,$\approx$\,6. These authors also showed that the diffuse low density interstellar medium exposed to UV radiation has a different relation to the extinction as compared to the dense cold ISM. Recently, \citet{Munari+2008} presented 68 measurements of  EW8620\,\AA~in the spectra from the RAVE survey ($R \sim 7500$) obtaining the relation $\displaystyle E_{B-V} = (2.72 \pm 0.03) \times EW8620$\,\AA~for $E_{B-V} < 1.2$. In this work we extend that relation to higher extinction values, which is relevant for the innermost Galactic region of the GAIA survey. 
 
\begin{figure}
\centering
\resizebox{\hsize}{!}{\includegraphics[width=20.0cm,angle=0, trim= 0cm 0.5cm 0.5cm 2cm, clip]{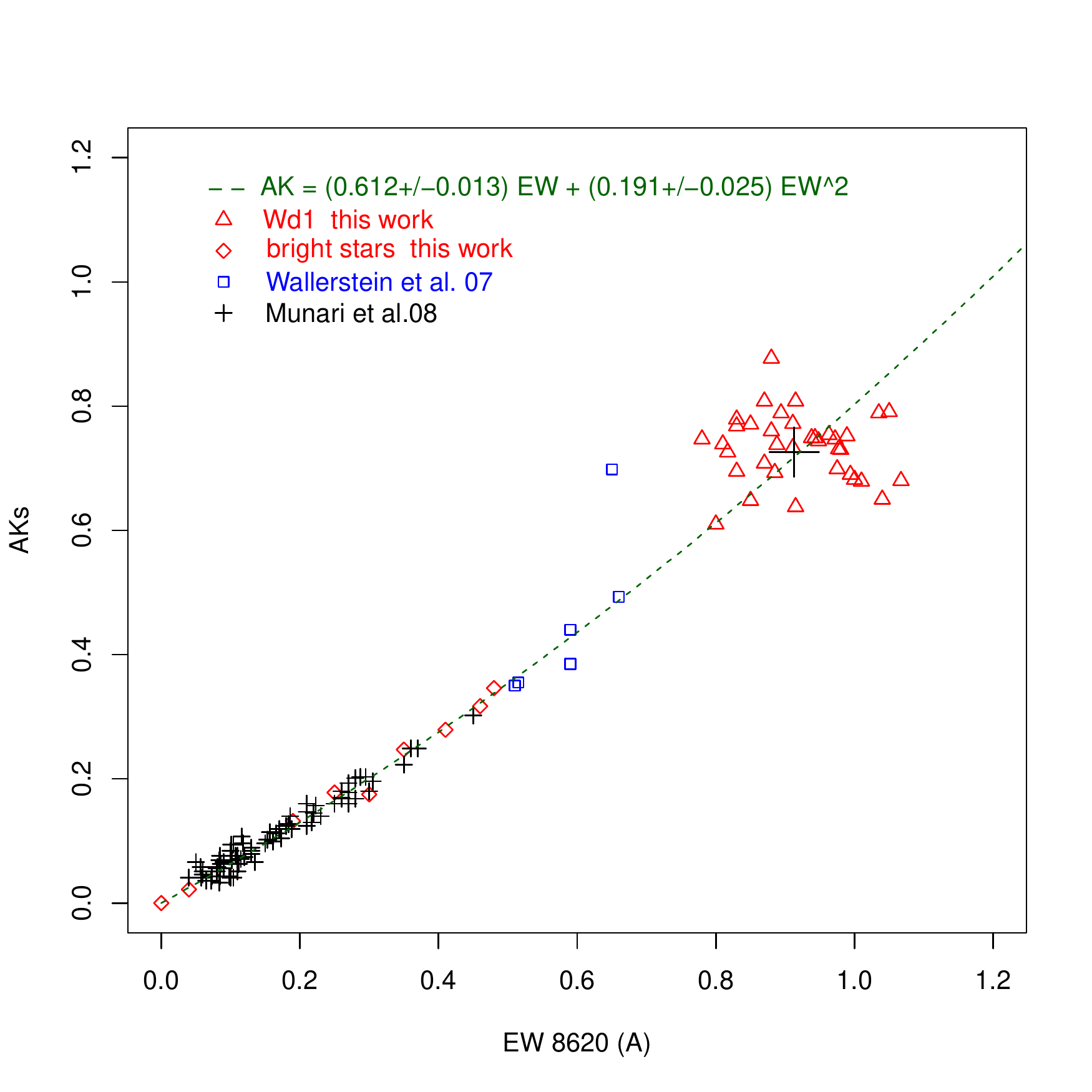}}
\caption{$A_{Ks}$ extinction $versus$ EW8620\,\AA.  Red symbols: this work -- triangles Wd~1 members, diamonds bright field stars. Black crosses \citep{Munari+2008}. Blue squares: \citet{Wallerstein+2007}. For EW8620\,$<$\,0.6\,\AA which translates to $E_{B-V}$\,$<$\,1 or $A_V$\,$<$\,7, the relation is linear and coincides with that of \citet{Munari+2008}. The size of the black cross indicates the 1-$\sigma$ standard deviation for the Wd~1 cluster members.}
\label{fig:AKEW8620}
\end{figure}

Spectra were measured as described in Sect.\,\ref{section2.3}. For Wd~1 cluster members, $A_{Ks}$ were measured in the present work. For the Munari and Wallerstein stars, we used the 2MASS photometry and spectral type taken from {\tt SIMBAD}\footnote{http://simbad.u-strasbg.fr/} to derive the extinction. Table 2 displays our results of EW8620\,\angstrom (column 2), $A_{Ks}$ (column 3) and the source for the 8620\,\AA\ data (column 4). We adopt $A_{Ks}$ instead of $A_V$ or $E_{B-V}$ because it is much less sensitive to dust size than optical wavelengths, and because our photometry was done in $JHKs$  bands. In this way, we use the relation we obtained for the Wd~1 cluster members: $A_{Ks}$\,=\,0.29\,$E_{B-V}$ to transform the extinction reported by \citet{Munari+2008}, \citet[][]{Wallerstein+2007} and for the nine field OB stars without $A_{Ks}$ measurements. The linear relation obtained by \citet[][]{Munari+2008} was thus transformed to $A_{Ks}$\,=\,0.691\,$\times$\,$EW8620$\,\AA. Combining these values with those derived here, we fit the set with a polynomial function. The relation we derive from the combined dataset is:

\begin{table}
\label{table2}
\tiny
\centering
\caption{EW8620\,\AA\,, $A_{Ks}$  and $A_V$ extinction. Sources identified as Schulte\# are reported in \citet{Wallerstein+2007}. First 10 rows of the table are presented. A full version containing 103 entries is available online at the CDS.}
\begin{tabular}{lllll}
\hline
 Identification &EW8620 (\AA)& $A_{Ks}$& $A_{V} $     & Spec. source  \\
\hline
\hline
W2a         &  0.885  & 0.686  &10.068& OPD\\
W6a         & 1.067   & 0.649  &10.985& OPD  \\
W6b         & 0.817   & 0.719  &10.970& 081.D-0324  \\
W7          & 0.844   & 0.772  &11.773& OPD  \\
W8a         & -       & 0.641  &11.033& OPD  \\
W8b         & 0.830   & 0.760  &10.411& 081.D-0324  \\
W11         &    -    & 0.734  &10.765&  \\
W12a        & 0.870   & 0.800  &12.543&  \\
W13         & 1.05    & 0.784  &10.818& 081.D-0324  \\
W15         & 0.948   & 0.733  &12.258& 081.D-0324  \\
\hline
\end{tabular}
\end{table}

\begin{equation}
A_{Ks} = (0.612 \pm 0.013)~EW + (0.191 \pm 0.025)~EW^2    
\label{eq25}
\end{equation}

This equation is in excellent agreement with \citet{Munari+2008} which is linear for $EW$\,$<$\,0.6\,\AA (equivalent to $A_V$\,$<$\,9 for the inner GP extinction law of this paper). 

\section{Discussion and Conclusions}\label{section7}

We present a study of the interstellar extinction in a FOV 10$'$ $\times$ 10$'$ in direction of the young cluster Westerlund~1 in $JHKs$ with photometric completeness $>$\,90\% at $\,$Ks$\,$=$\,15$. Using data publicly available, we extended the wavelength coverage to shorter and longer wavelengths from the optical to the MIR (although with less complete  photometry). Colour excess ratios were derived by combining (92) Wd~1 cluster members with published spectral classification with (8463) RC stars inside the FOV. 

Our result for the NIR: $E_{J-H}/E_{H-Ks} = 1.891\pm0.001$ is typical of recent deep imaging towards the inner Galaxy. Using the procedure designed by \citet{Stead+2009} to obtain effective wavelengths of 2MASS survey and Eq.\,\eqref{eq2} we derived a power law exponent $\alpha = 2.126 \pm 0.080$ that implies $A_J/A_{Ks} = 3.23$. This extinction law is steeper than the older ones \citep{Indebetouw+2005,CCM89,RL85}, based on less deep imaging and is in line with recent results based on deep imaging surveys \citep{Stead+2009, Nishiyama+2006, Nataf16}. In the NIR, this implies in smaller $A_{Ks}$ and larger distances than laws based on more shallow photometry, which has a large impact on inner Galaxy studies.

Using our measured $A_{Ks} / E_{J-Ks} = 0.449$ (plus combinations between other filters) we obtained the extinction to Wd~1, $<A_{Ks}> = 0.736 \pm 0.056$. This is $0.2-0.3$ magnitudes smaller than previous work, based on older (shallower) extinction laws \citep{Gennaro+2011, Negueruela+2010, Lim+2013, Piatti+1998}. On the other hand our $A_V = 11.26 \pm 0.07$ is in excellent agreement with $A_V = 11.6$ derived by \citet{Clark+2005} based on a completely different method: the OI~7774\,\AA\,EW $\times$ $M_V$ of six Yellow Hypergiants. 
The cluster extinction encompass the range $A_{Ks} = 0.55-1.17$ (which translates into $A_{V} \approx 8.5-17$). Cluster members have typical extinction $A_{Ks}=0.74\pm 0.08$ which translates into $A_V=11.4\pm 1.2$. The foreground interstellar component is $A_{Ks} = 0.63\pm 0.02$  or $A_V = 9.66\pm 0.30$.

The extinction spread of $A_V \sim 2.5$ magnitudes inside a FOV 3.5$\arcmin$ indicates that it  is produced by dust connected to the cluster region. In fact Fig\ref{fig:dustmap} shows a patchy distribution of warm dust.  There are indications for a gradient in $A_{Ks}$ increasing from SW to NE, which is in line with the map of warm dust  and with the colour density maps in the surrounding field. However, the effect is not very clear, suggesting a patchy intra-cluster extinction. The $J-Ks$ colour density maps unveiled the existence of a group of blue foreground stars, which may or may not be a real cluster. Since those stars partially overlap the Wd~1 cluster, they must be taken into account when subtracting the field population in the usual procedures to isolate Wd~1 cluster members.

We measured the EW8620\,\AA\,DIB for 43 Wd~1 cluster members and combined them with additional filed stars and results collected from the literature, showing a good correlation with $A_{Ks}$. Although the linear relation reported by \citet{Munari+2008} was recovered for $E_{B-V}$\,$\approx 1$, it deviates for larger values and we present a polynomial fit extending the relation. The moderately large scatter in the Wd~1 measurements seems to reflect the uncertainties in our procedures to measure the extinction and the EW. Unfortunately our sample does not probe the range 0.4\,$<$\,EW8620\,$<$\,0.8, see Fig.\,\ref{fig:AKEW8620}. In order to improve the situation aiming for GAIA, the above relation should be re--done incorporating measurements from stars with 1$\,<$\,$E_{B-V}$\,$<$\,2.5 (equivalent to 0.3\,$<$\,$A_{Ks}$\,$<$\,0.7). As a matter of fact, it is expected that such relation will be different for the general ISM as compared to the denser ambients prevalent inside regions of recent star forming regions.

We examined our result $R_V\,=\,A_V /E_{B-V}\,=\,2.50 \pm 0.04$ with care, since previous works found suspicious that this ratio for Wd~1 was much smaller than the usual $R_V$\,=\,3.1 \citet{Clark+2005,Lim+2013, Negueruela+2010}, although similar values have been reported by \citet{FM09} for a couple of stars like star BD+45~973. Moreover, this value is in excellent agreement with \citet{Nataf13}, as deduced from  $R_I/E_{V-I}$ based on OGLE~III fields in the Galactic Bulge, close to the position of Wd~1. However, even if there was minor systematic errors in the photometric calibrations in the B--band, it would not be larger than a few tenths of magnitudes, which are dwarfed by the large extinctions:  $A_B \approx 15$. 

An interesting result is the lack of correlation between the reddening law in the optical, as compared with the NIR (see Fig.\ref{fig:RJKIJxRAIVI}). This is in agreement with the result found by \citet{Nataf16} for a much larger field in the inner Galaxy. Looking to the position of \citet{CCM89} in that figure, we confirm other diagnostics, showing that dust grain properties in the inner Galaxy are different from those sampled by shallower imaging.  Even in our small field (12\arcmin$\times$12\arcmin), the colour ratio diagram of Fig.\,\ref{fig:RJKIJxRAIVI} shows that the Wd~1 cluster members spread to a different zone in the diagram, as compared to RCs. This suggests that intra--cluster dust grains have properties different from the lower density ISM where RCs are located. The large spread of indices between Wd~1 members indicates the existence of clumps of dust grains with a variety of size properties. 

We derived the extinction law for the range 0.4-4.8\,$\mu$m which is in very good agreement with the colour excess ratios obtained by \citet{Scha16} from large photometric and spectroscopic surveys in the inner GP. We propose our law presented in Eq.\,\eqref{eq19} to be representative of the average inner GP. A striking feature of this law is its close coincidence with a power law  with exponent $\alpha = 2.17$ for the entire range 0.8-4\,$\mu$m. We call the reader's attention to the fact that this is an average law, usefull for general purposes, since the reddening law varies from place to place inside the narrow zone of the GP. 

\section*{Acknowledgements} 
We thank an anonymous referee for the very productive questions/remarks which has improved our manuscript. We also thank  M. Gennaro and A. Bonanos for a critical comments on earlier versions of this paper. AD thanks to Funda\c{c}\~{a}o de Amparo {\'a} Pesquisa do Estado de S\~{a}o Paulo - FAPESP for support through proc. 2011/51680-6. LAA acknowledges support from the FAPESP (2013/18245-0 and 2012/09716-6). FN acknowledges support from the FAPESP (2013/11680-2). This research has made use of the VizieR catalogue access tool, CDS, Strasbourg, France.

\appendix
\section{Effective wavelengths for 2MASS filters}\label{appendixA}

\citep{Stead+2009} obtained the effective wavelengths for the JHKs filters of 2MASS system by convolving the filter bandpasses with model atmospheres reddened by different amounts of extinction (0 to 3 in Ks band) and their results are in their Fig. 2a. Since the transformation curves approximately straight lines, we found useful to have linear expressions to obtain the effective wavelengths from the isophotal ones. With this approach, it easy to obtain accurate values for the exponent $\alpha$ of the reddening power law based just on the E(H-K) colour excess. This can be applied for every single star which has a known intrinsic colour.

\begin{eqnarray}
\lambda_{Jeff}     & = & 1.235~[1 + 0.0214~E(H-K)]	\label{A1}	\\
\lambda_{Heff}     & = & 1.662~[1 + 0.0104~E(H-K)]	\label{A2}	\\
\lambda_{Jeff}     & = & 2.159~[1 + 0.00415~E(H-K)]	\label{A3}
\end{eqnarray}

\section{Plots of measured colour excess ratios}\label{AppendixC}

\begin{figure*}
\begin{minipage}{176mm}
\label{fig:appA}
\centering
\subfloat{\includegraphics[width=5.6cm]{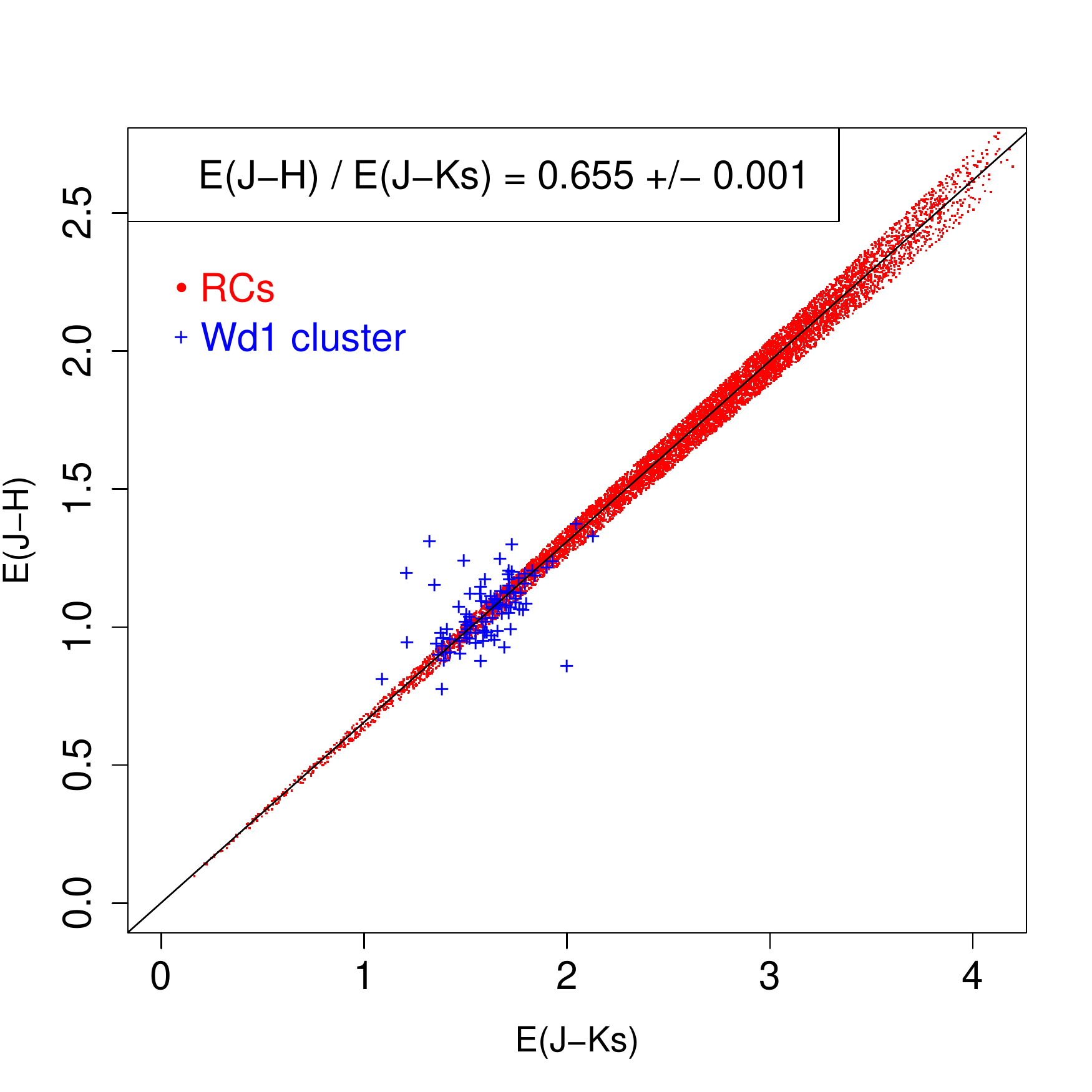}}\,
\subfloat{\includegraphics[width=5.6cm]{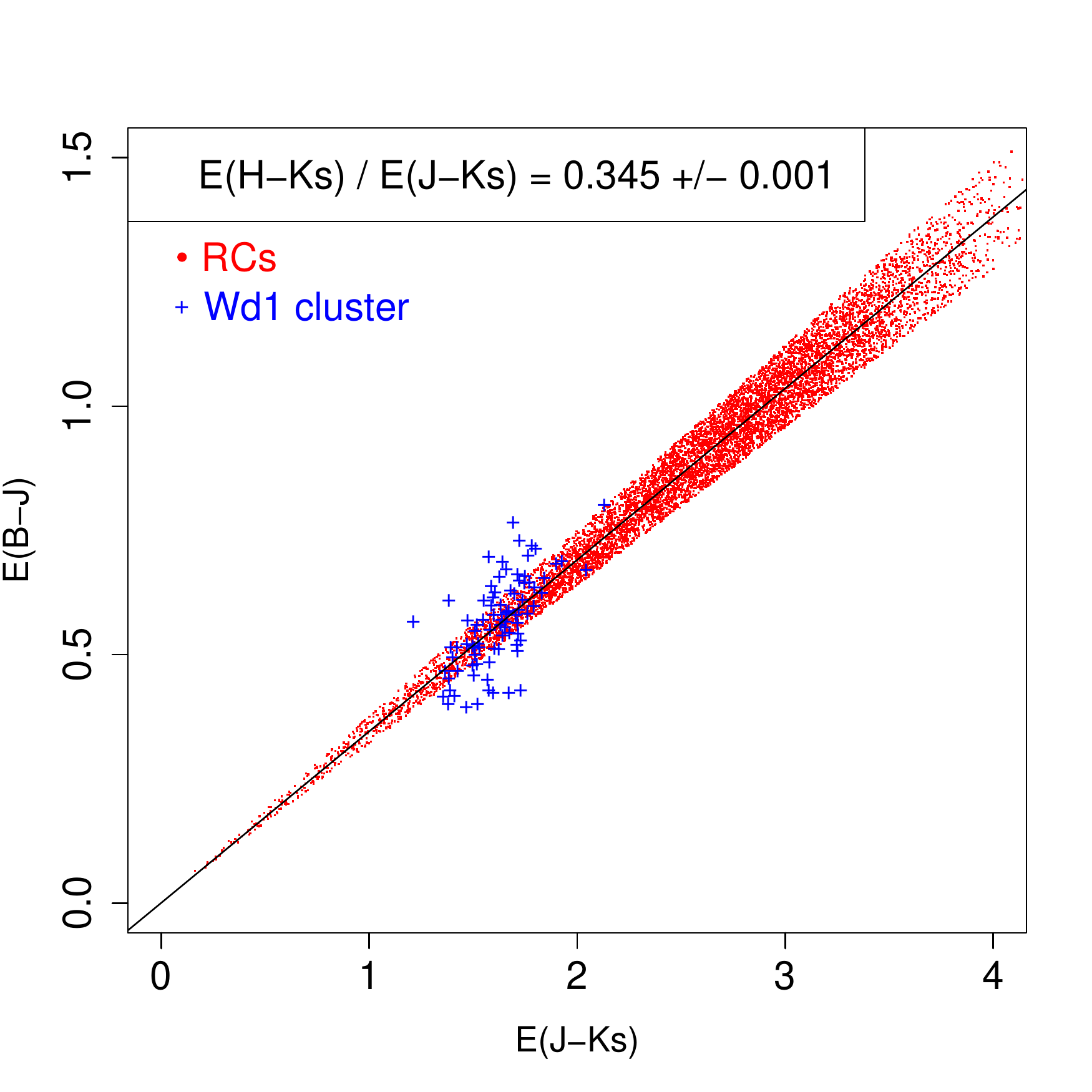}}\,
\subfloat{\includegraphics[width=5.6cm]{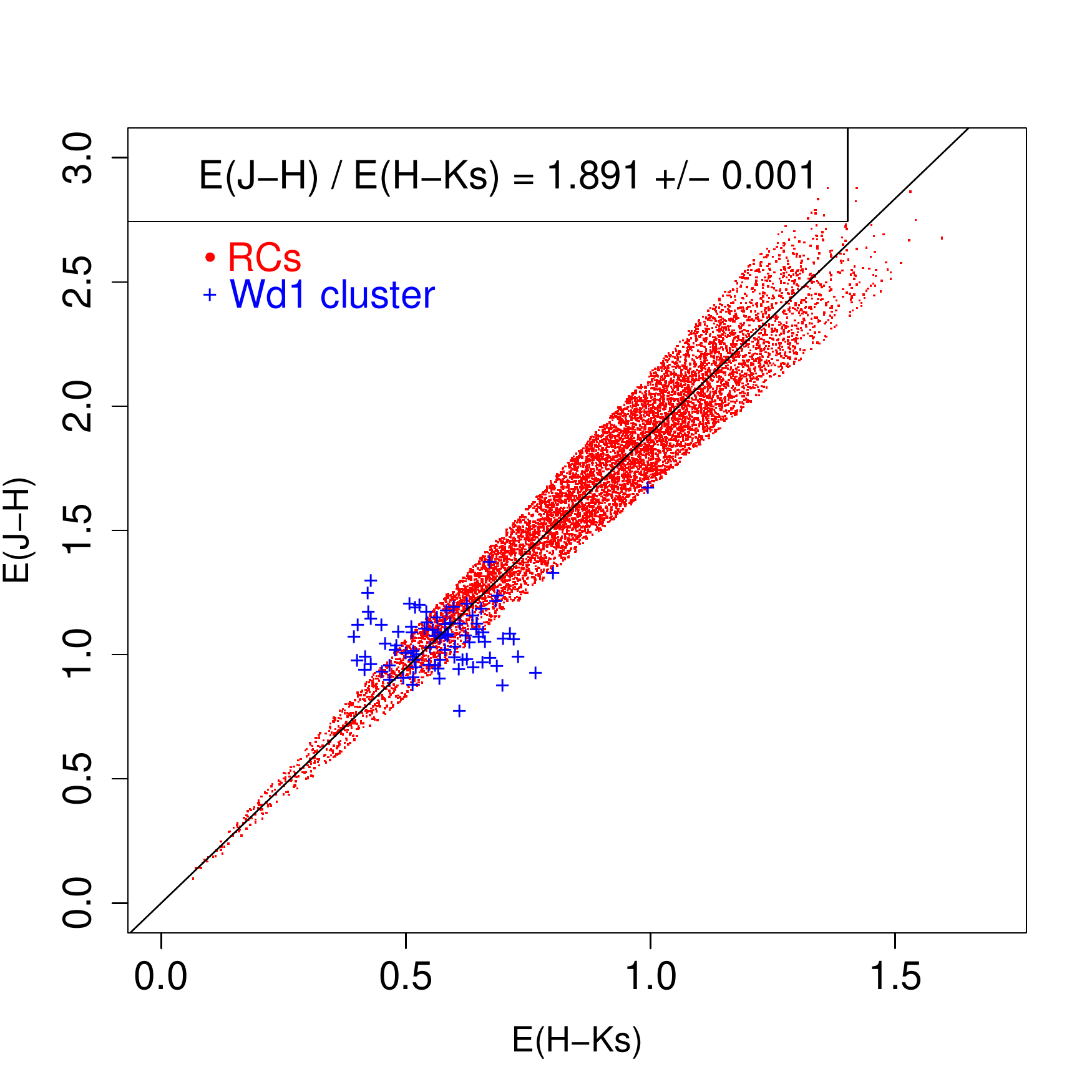}}\,
\caption{Colour excess ratios involving $JHKs$ filters. Both samples of RCs and Wd~1 cluster were used together to perform the linear fit.}
\end{minipage}
\end{figure*}

\begin{figure*}
\begin{minipage}{176mm}
\label{fig:appB}
\centering
\subfloat{\includegraphics[width=5.6cm]{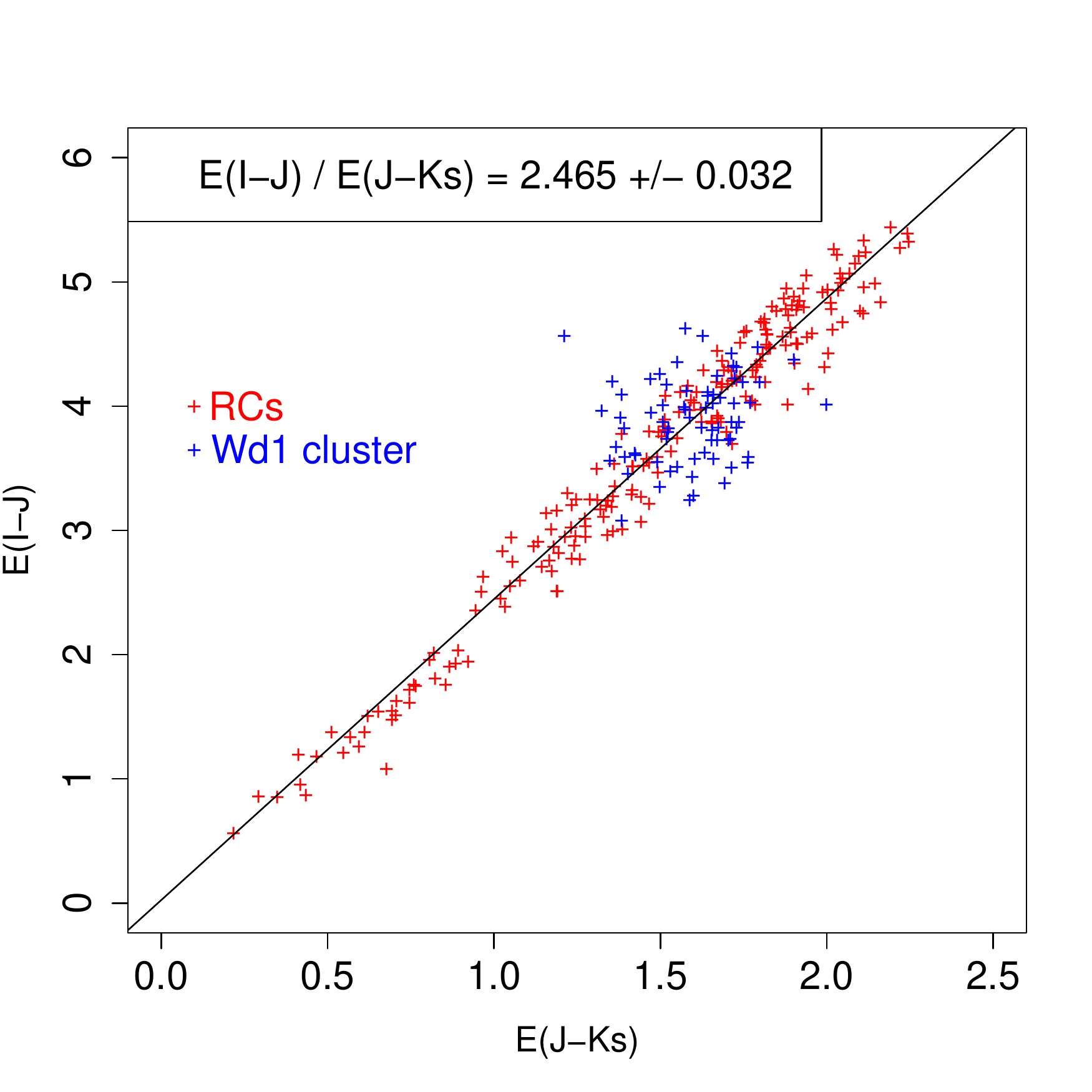}}\,
\subfloat{\includegraphics[width=5.6cm]{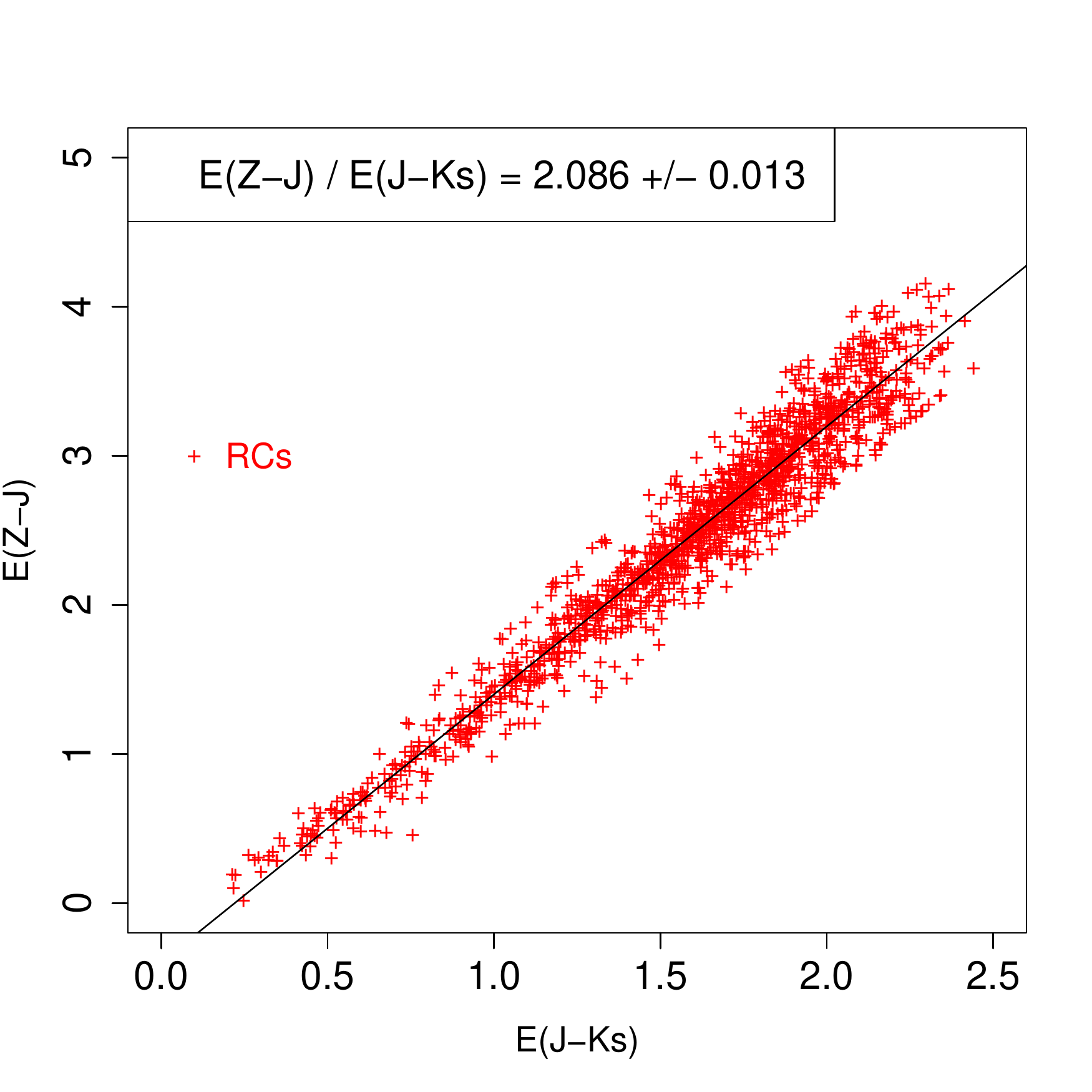}}\,
\subfloat{\includegraphics[width=5.6cm]{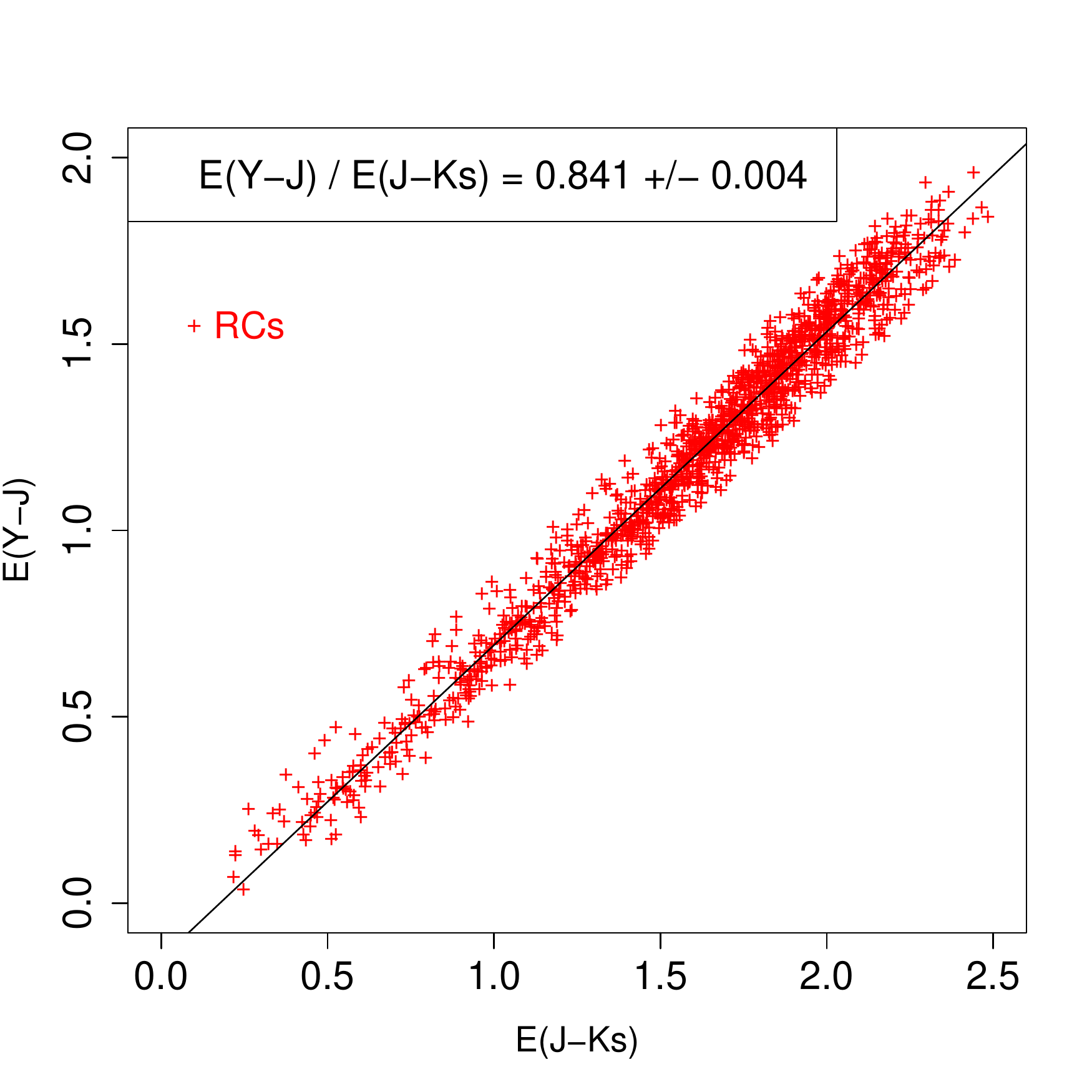}}\,
\caption{Similar to Fig.\,\ref{fig:appA}.}
\end{minipage}
\end{figure*}

\begin{figure*}
\label{fig:appC}
\begin{minipage}{176mm}
\centering
\subfloat{\includegraphics[width=5.6cm]{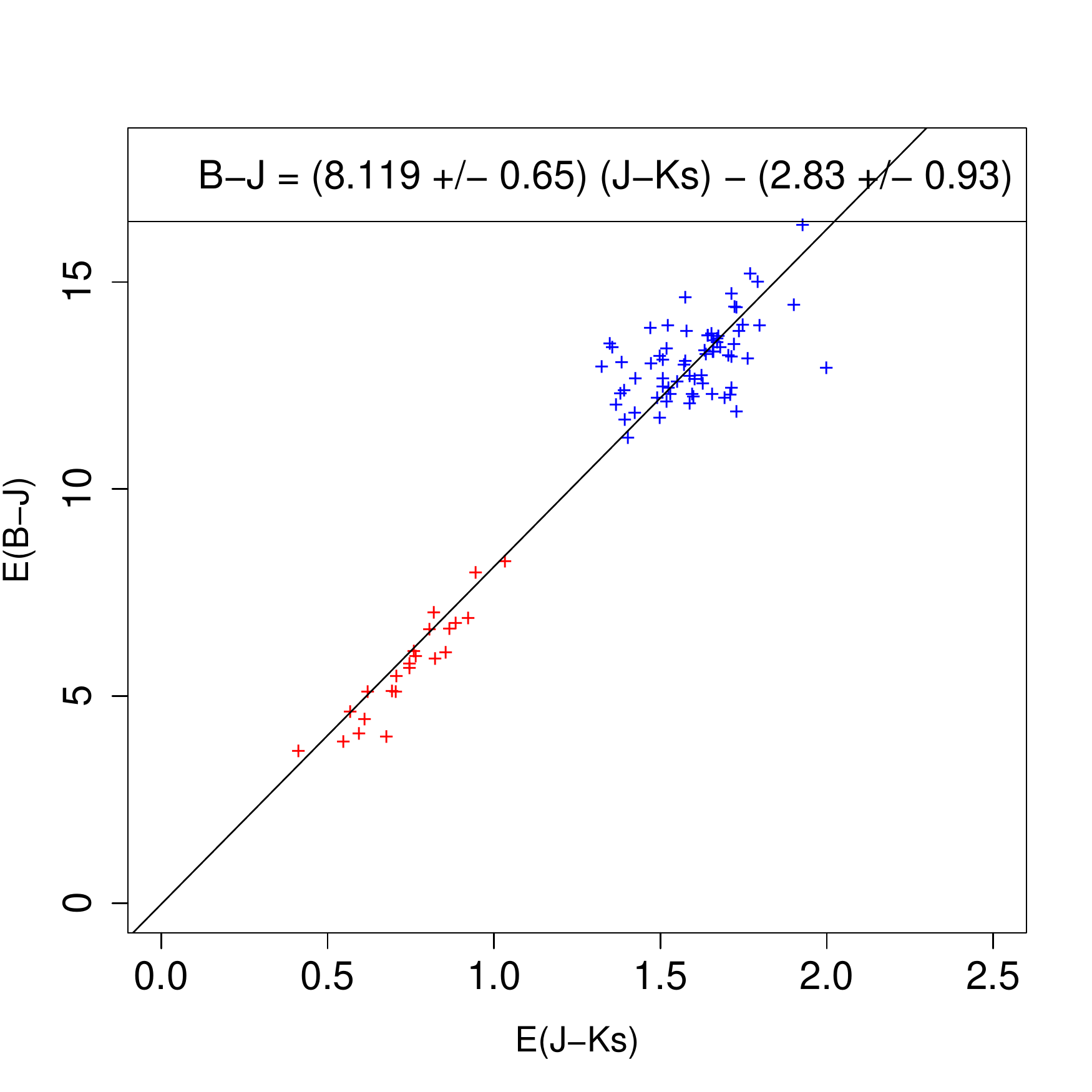}}\,
\subfloat{\includegraphics[width=5.6cm]{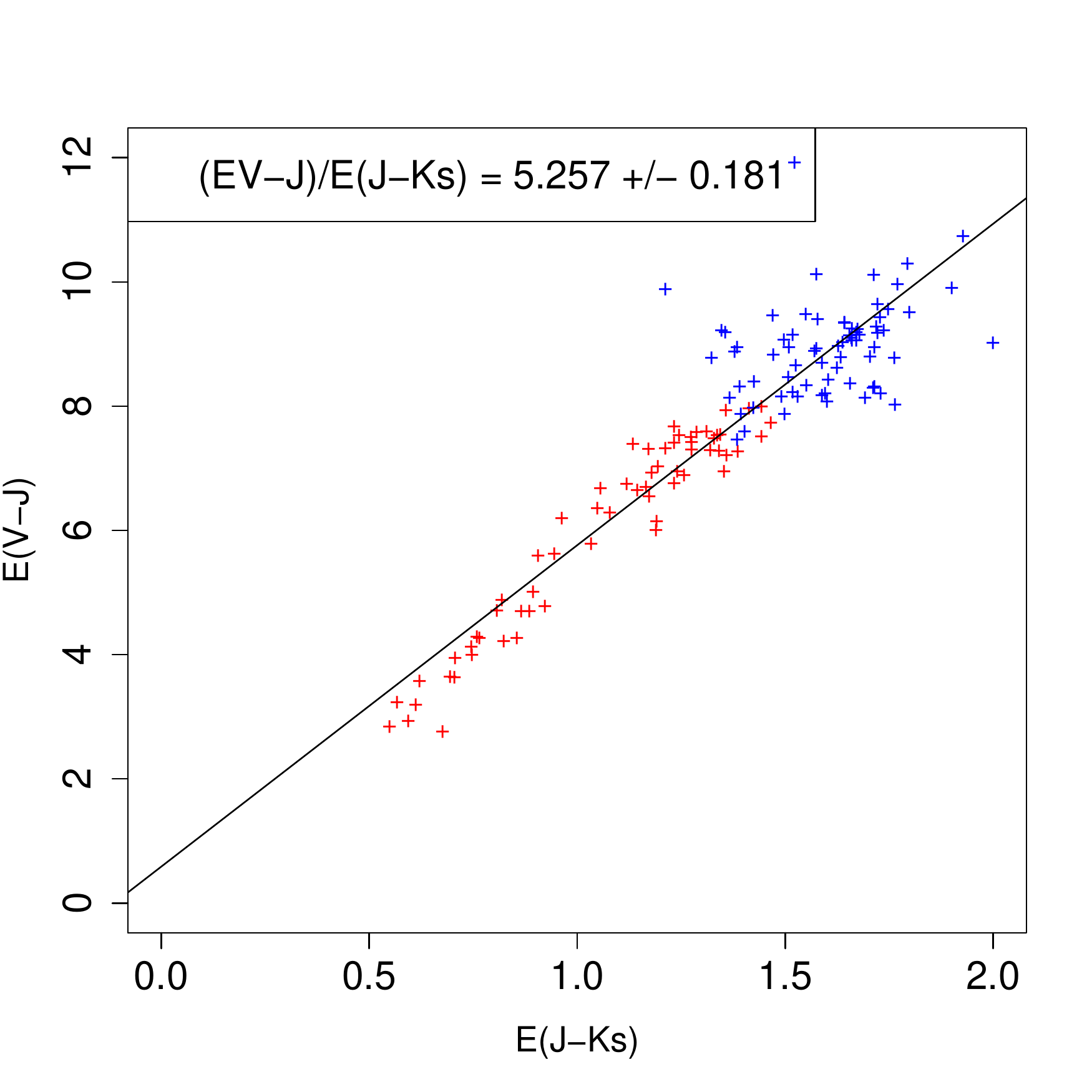}}\,
\subfloat{\includegraphics[width=5.6cm]{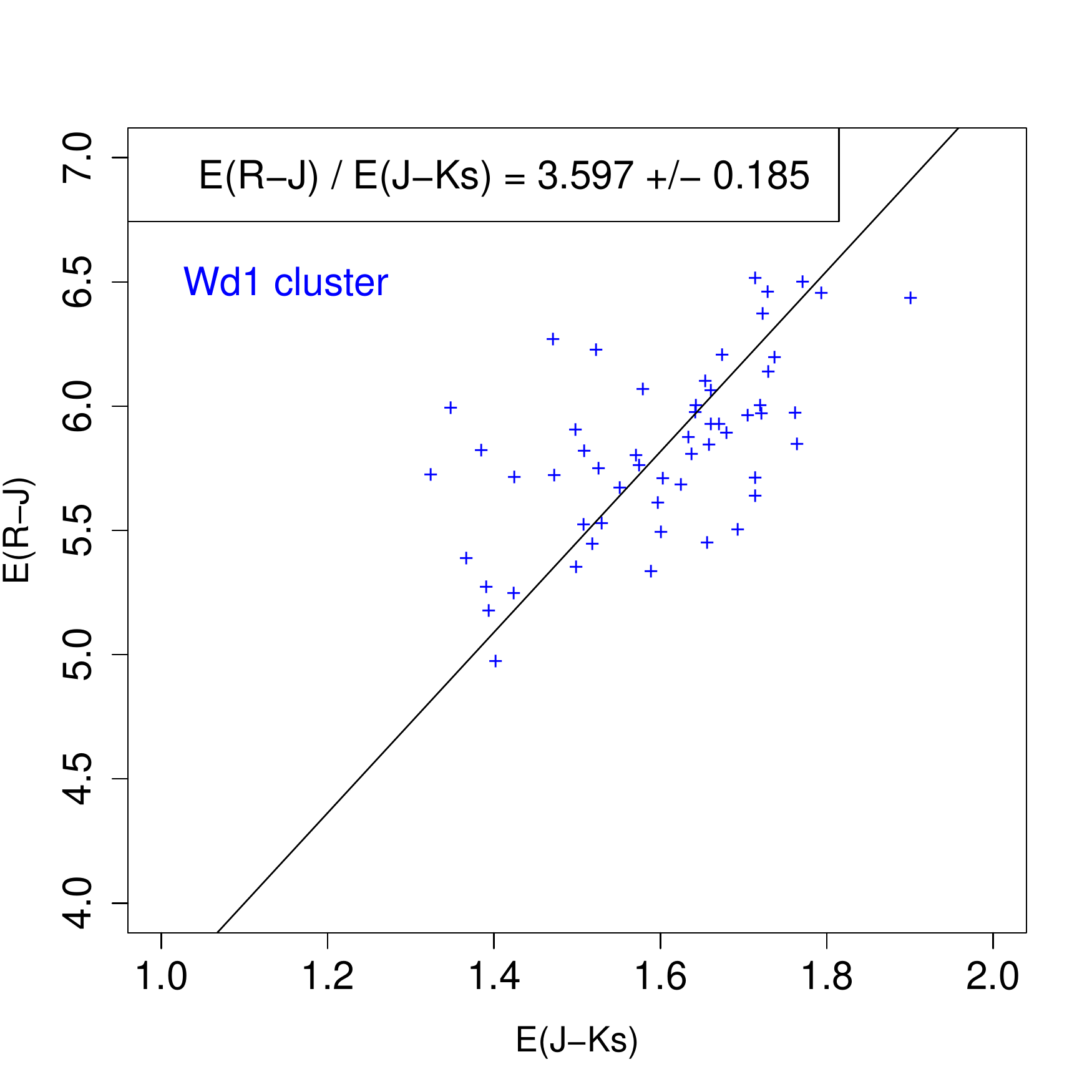}}\,
\caption{Similar to Fig.\,\ref{fig:appA}.}
\end{minipage}
\end{figure*}

\begin{figure*}
\label{fig:appD}
\begin{minipage}{176mm}
\centering
\subfloat{\includegraphics[width=5.6cm]{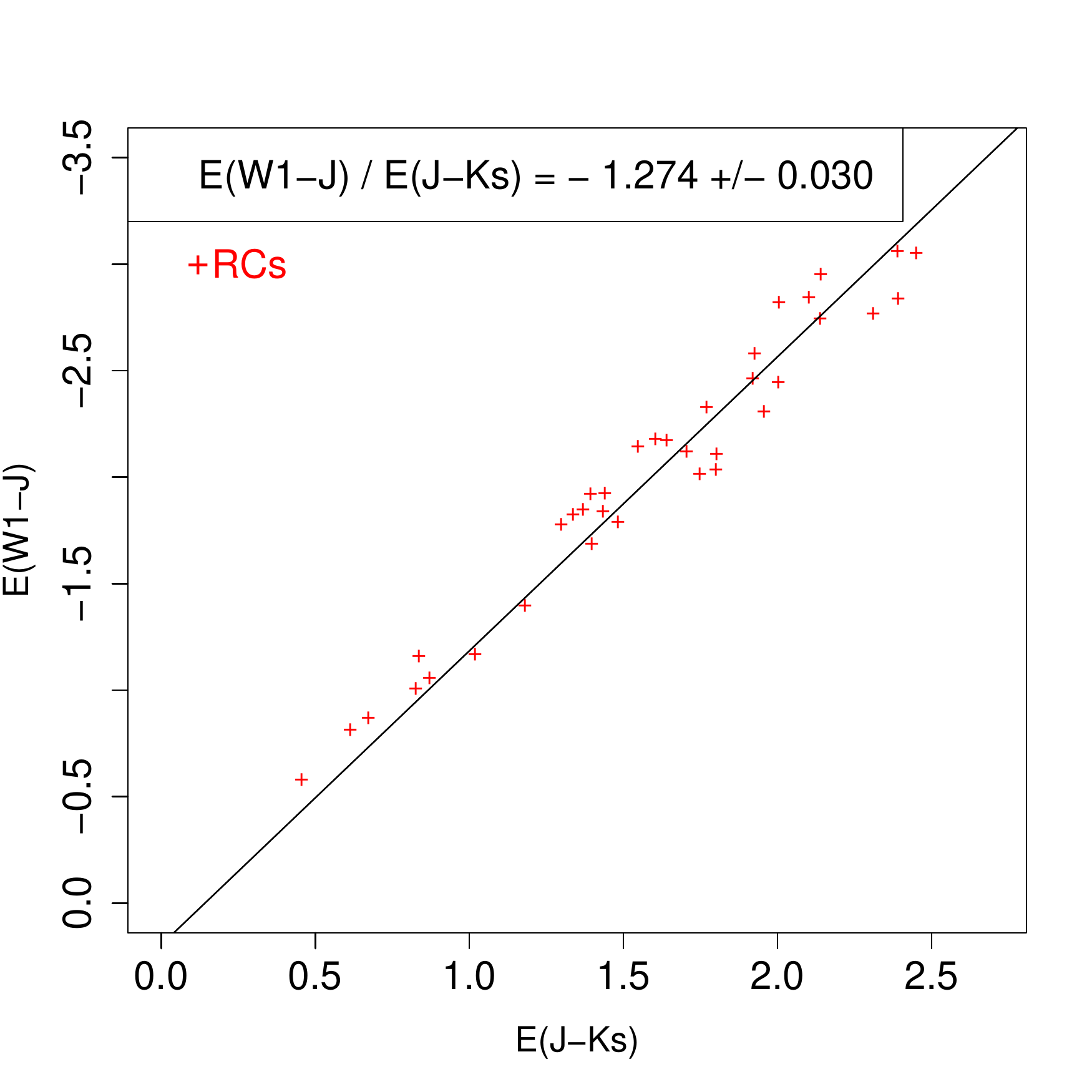}}\,
\subfloat{\includegraphics[width=5.6cm]{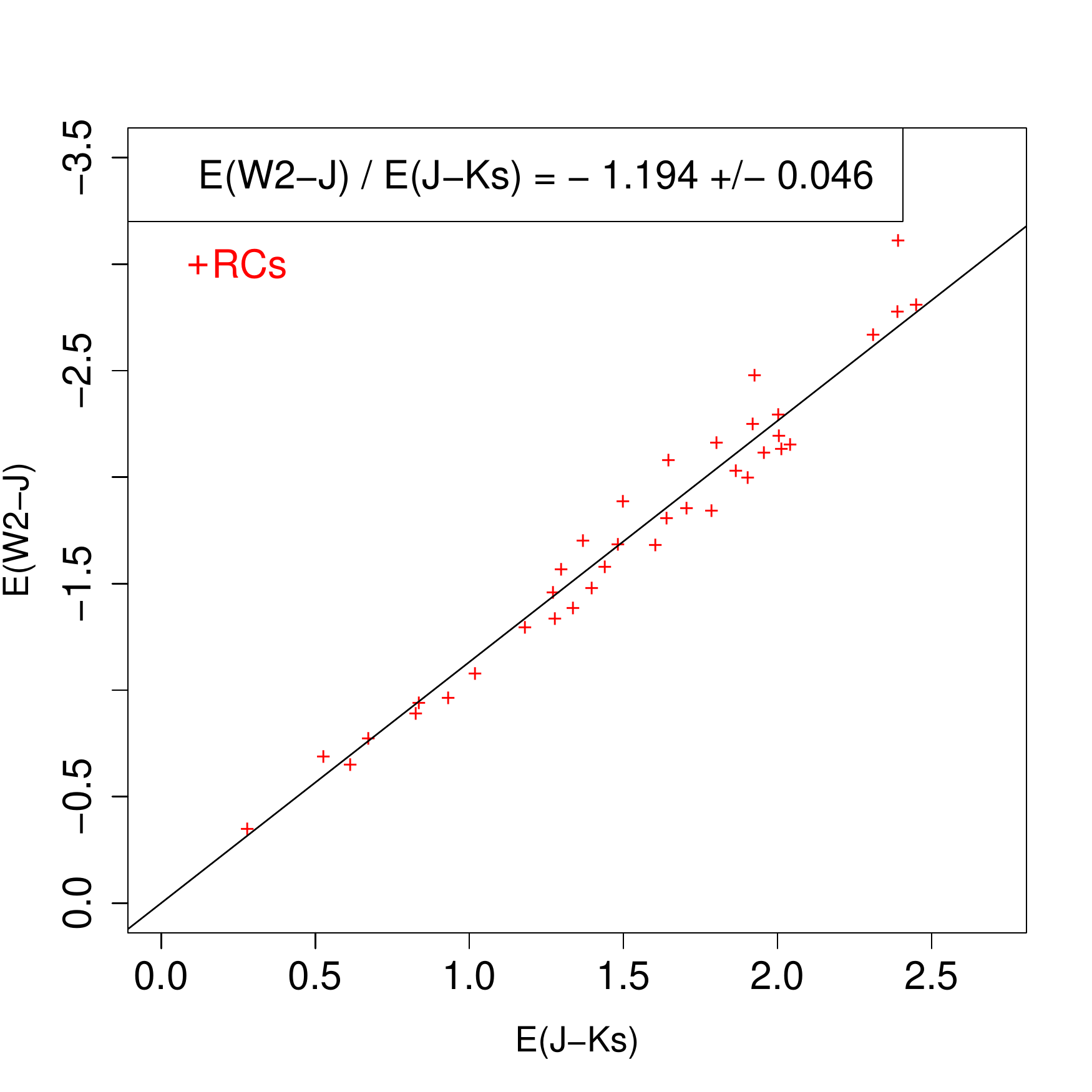}}\,
\subfloat{\includegraphics[width=5.6cm]{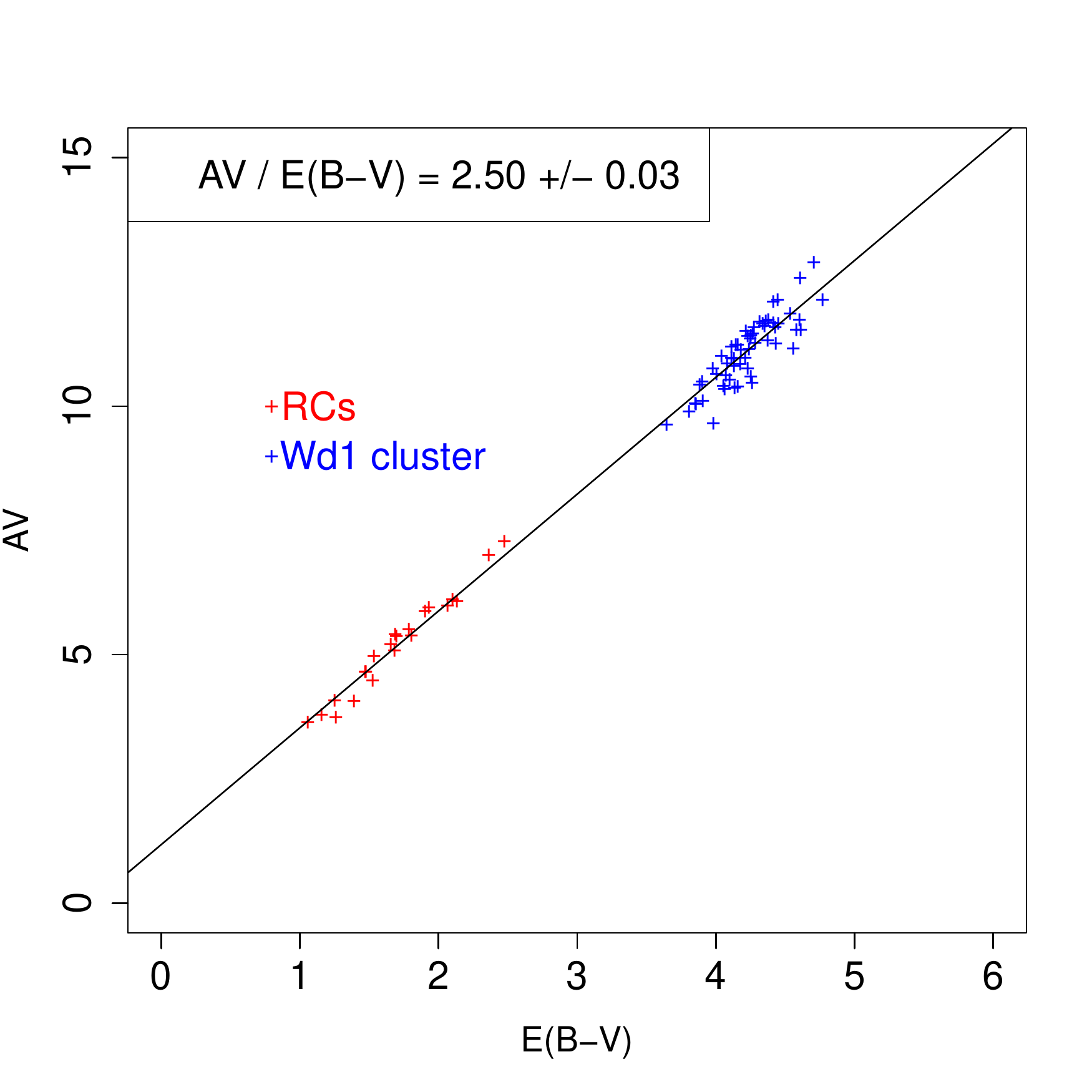}}\,
\caption{Similar to Fig.\,\ref{fig:appA}.}
\end{minipage}
\end{figure*}

\end{document}